\documentclass[aps,notitlepage,twocolumn,showpacs,preprintnumbers,floatfix,tightenlines, longbibliography, nofootinbib]{revtex4-1}
\usepackage{mathrsfs}
\usepackage{amssymb}
\usepackage{amsmath}
\usepackage{graphicx}
\usepackage{amsfonts,amsbsy}
\usepackage{nicefrac}
\usepackage{slashed}
\usepackage{xcolor}
\usepackage[breaklinks,colorlinks,urlcolor=blue,citecolor=citcolor,linkcolor=lcolor]{hyperref}
\usepackage{array}
\usepackage{multirow}
\usepackage{siunitx}
\usepackage[normalem]{ulem} %normalem disables the feature that ulem automatically renders \emph{} to \underline{}.

\definecolor{lcolor}{rgb}{0.5,0,0}
\definecolor{citcolor}{rgb}{0,0.3,0.0}
\definecolor{ao(english)}{rgb}{0.0, 0.5, 0.0}
\definecolor{RoyalBlue}{HTML}{0071BC} 
\usepackage{comment}

\newcommand{\mrm}{\mathrm}

\newcommand{\dA}{d_{\mrm{A}}}

\newcommand{\NC}{N_\mathrm{c}}
\newcommand{\fig}{Fig.~}

\newcommand{\eq}{Eq.~}
\newcommand{\se}{Sec.~}

\newcommand{\re}{Ref.~}

\newcommand{\tab}{Table~}
\newcommand{\app}{Appendix~}

\newcommand{\qperp}{q_\perp}

\newcommand{\lperp}{\Lambda_\perp} 
\newcommand{\qhat}{\hat q}

\newcommand{\vb}{\vec}
\renewcommand{\vec}[1]{\mathrm{\mathbf{#1}}}
\newcommand{\dd}[2][]{\mathrm d^{#1}{#2}\,} 

\newcommand{\pdv}[2][]{\frac{\partial{#1}}{\partial{#2}}}

\newcommand{\pmin}{p_{\mathrm{min}}}
\newcommand{\pmax}{p_{\mathrm{max}}}

\newcommand{\taubmss}{\tau_{\mathrm{BMSS}}}
\newcommand{\tauR}{\tau_R}
\newcommand{\tautherm}{\tau_{\mathrm{therm}}}
\newcommand{\ttherm}{t_{\mathrm{therm.}}}
\newcommand{\tauhydro}{\tau_{\mathrm{hydro}}}
\newcommand{\Conetwo}{\mathcal {C}^{1\leftrightarrow 2}}
\newcommand{\Ctwotwo}{\mathcal{ C}^{2\leftrightarrow 2}}

\newcommand{\Teps}{T_{\varepsilon}}

\newcommand{\thetaqp}{\theta_{qp}}
\newcommand{\thetaqk}{\theta_{kq}}
\newcommand{\phiqk}{\phi_{qk}}
\newcommand{\phiqp}{\phi_{qp}}

\newcommand{\Tr}{\mathrm{Tr}\,}

\newcommand{\PiLR}{{\Pi^{00}}}

\newcommand{\PiTR}{{\Pi^{T}}}
\newcommand{\GLR}{{G^{00}}}
\newcommand{\GTR}{{G^{T}}}
\newcommand{\RE}{\mathrm{Re}}
\newcommand{\IM}{\mathrm{Im}}

\newcommand{\Tid}{T_{\mathrm{id}}}
\newcommand{\Tfirst}{T_{\mathrm{1st}}}

\newcommand{\Gret}{G}
\newcommand{\GretHTL}{G^{\mathrm{isoHTL}}}
\newcommand{\epsgoal}{\epsilon^{\mathrm{error goal}}}

\newcommand{\xigauge}{\xi}
\newcommand{\xiscreen}{{\xi_g}}
\newcommand{\xiscreensqr}{{\xi_g^2}}
\newcommand{\xianiso}{\xi_0}

\newcommand{\Mhtl}{\mathcal M_{\mathrm{HTL}}}
\newcommand{\Mdebyeone}{\mathcal M_{\mathrm{Debye1}}}
\newcommand{\Mdebyetwo}{\mathcal M_{\mathrm{Debye2}}}
\newcommand{\Mdebyethree}{\mathcal M_{\mathrm{Debye3}}}

\newcommand{\vbphat}{\hat{\vb p}} %unit vector in direction of p

\DeclareSIUnit\c{c}

\begin{document}

\title{Soft-gluon exchange matters: isotropic screening in QCD kinetic theory}

\author{K.~Boguslavski} 
\author{F.~Lindenbauer} 
\email{florian.lindenbauer@tuwien.ac.at}
\affiliation{Institute for Theoretical Physics, TU Wien, Wiedner Hauptstraße 8-10, 1040 Vienna,
Austria}

\begin{abstract}
    QCD kinetic theory simulations are a prominent tool for studying the nonequilibrium initial stages in heavy-ion collisions. Despite their success, all implementations rely on approximations of the hard thermal loop (HTL) screened matrix elements in the collision terms. In this paper, we present our results for different isotropic screening prescriptions in the elastic collision term for gluons. In particular, we go beyond the simple Debye-like screening form that is used in all current implementations and apply the isotropic HTL matrix element instead. For isotropic systems, the evolution is nearly unchanged, but when studying the equilibration process in a Bjorken expanding plasma, we find qualitative and quantitative differences in a range of moments of the distribution function, such as a decrease in the maximum pressure anisotropy by up to 50\%. In contrast, we find no significant qualitative impact on the jet quenching parameter or 
    on the late-time hydrodynamization dynamics of anisotropic plasmas but observe systematically smaller values of $\eta/s$ by about $10 \%$ that coincide with perturbative calculations at small couplings.
    Our study reveals that the choice of the screening prescription can lead to large corrections at early times, but is less important close to equilibrium.
\end{abstract}

\maketitle

\section{Introduction}
Numerical QCD kinetic theory simulations are now a state-of-the-art tool to simulate the initial nonequilibrium stages in heavy-ion collisions. They are shown to numerically reproduce the bottom-up thermalization process \cite{Baier:2000sb, Kurkela:2015qoa}, 
smoothly connect to the later hydrodynamic evolution \cite{Kurkela:2015qoa, Kurkela:2018xxd, Kurkela:2018oqw, Kurkela:2018wud, Kurkela:2018vqr, Du:2020dvp, Du:2022bel, Boguslavski:2023jvg} and are also used to study transport coefficients during the initial stages \cite{Boguslavski:2023alu, Boguslavski:2023fdm, Boguslavski:2023waw, Du:2023izb}. However, although QCD effective kinetic theory (EKT) was originally formulated over 20 years ago \cite{Arnold:2002zm}, 
current implementations \cite{Kurkela:2015qoa, Kurkela:2018xxd, Du:2020zqg, kurkela_2023_10409474} have soft screening approximations the impact of which has not yet been studied systematically.

In principle, one should use the hard thermal loop (HTL) propagator\footnote{Out of equilibrium, the term \emph{hard loop} is often used. In this paper, we will use \emph{hard thermal loop} (HTL) also for out-of-equilibrium systems.} \cite{Braaten:1989mz} to account for medium effects in the collision integrals. However, for anisotropic plasmas as typically encountered during initial stages in heavy-ion collisions, this leads to the appearance of plasma instabilities \cite{Mrowczynski:1988dz, Arnold:2003rq, Romatschke:2003ms, Romatschke:2006bb, Kurkela:2011ub, Hauksson:2021okc}, which make a perturbative treatment problematic. Therefore, all current implementations employ an isotropic screening approximation with 
a simple Debye-like screening prescription \cite{AbraaoYork:2014hbk} that is used for the in-medium propagator in the elastic collision term.

In a previous paper \cite{Boguslavski:2023waw}, we have found for the jet quenching parameter $\qhat$ that a Debye-like screening prescription yields accurate values for large momentum cutoffs while showing larger deviations to HTL-screened values of $\qhat$ for smaller cutoffs. Additionally, isotropic Debye-like screening cannot accurately reproduce both longitudinal and transverse momentum diffusion simultaneously. This motivates us to study the effect of Debye-like screening approximations in comparison to HTL screening in dynamical kinetic theory simulations in this study.

In this paper, we, therefore, focus on the impact of different screening prescriptions in the elastic collision term on gluonic simulations of the bottom-up equilibration process. 
While this thermalization (hydrodynamization) process has been studied extensively in the literature \cite{Baier:2000sb, Kurkela:2014tea, Kurkela:2015qoa, Kurkela:2018xxd, Du:2020zqg, Brewer:2019oha, Fu:2021jhl, Brewer:2022vkq, BarreraCabodevila:2022jhi, Cabodevila:2023htm, Rajagopal:2024lou}, we include for the first time the full isotropic hard thermal loop (isoHTL) matrix element in a kinetic theory simulation. We compare its impact on the evolution of different observables to several forms of the often-used Debye-like screening prescriptions, which are approximations to the HTL matrix element using a simple screening mass \cite{AbraaoYork:2014hbk}. 
For isotropic plasmas, we find that the impact is negligible.

\begin{figure}
    \centering
    \includegraphics[width=\linewidth]{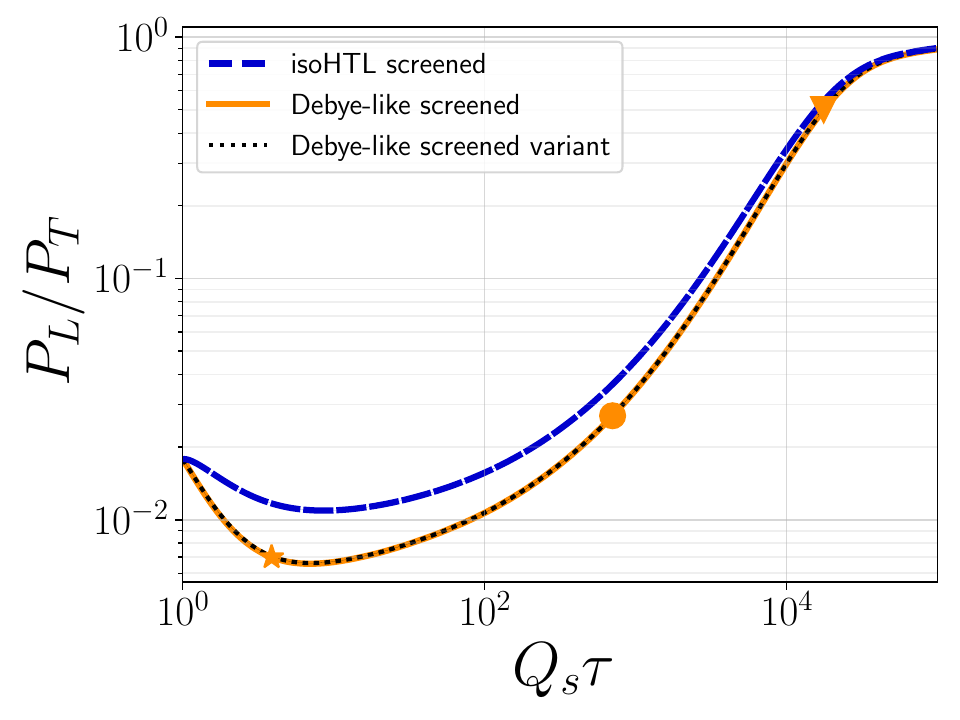}
    \caption{Pressure ratio of a longitudinally expanding plasma for $\lambda = 0.5$
    with different screening prescriptions. Both considered Debye-like screening prescriptions coincide while the isoHTL screened curve (dashed blue line) strongly deviates at early times, signaling less bulk anisotropy.
    }
    \label{fig:lambda05_runs}
\end{figure}

The main results of this paper 
are summarized in \fig \ref{fig:lambda05_runs} for simulations of the bottom-up equilibration process. At the example of 
the ratio of the longitudinal over transverse pressure at weak coupling,
we show its evolution for different screening prescriptions. 
We observe that the Debye-like screened curves (continuous orange and black dotted) lie on top of each other. At late times, the system approaches a hydrodynamic evolution in a similar way for all prescriptions.
This allows us to absorb most of the late-time modifications through isoHTL screening in a correction to the shear viscosity over entropy density transport coefficient $\eta/s$. In contrast, isoHTL screening leads to a significantly smaller pressure ratio at early times, almost by a factor of $2$. 
We find similar features for a wide range of observables, showing the importance of improved screening prescriptions for the far-from-equilibrium dynamics.

This paper is structured as follows: In Section~\ref{sec:methods}, we briefly review QCD kinetic theory and
how medium effects are included using the HTL propagator.
We introduce the precise form of the screening prescriptions that we study in section~\ref{sec:different_screening_prescriptions}, and provide our numerical results in section \ref{sec:thermalization_isotropic} for isotropic and in section \ref{sec:results} for expanding systems.
Finally, we conclude in Section \ref{sec:conclusions}. 
The Appendices include technical details on the discretization, kinematic considerations, form and validity of screening prescriptions, and the adaptive timestep employed in our studies.
Our data and analysis scripts can be found in \cite{lindenbauer_2024_13151024}.

\section{Effective kinetic theory\label{sec:methods}}
\subsection{Boltzmann equation}
\label{sec:Boltzmann}
We use QCD effective kinetic theory \cite{Arnold:2002zm} for a purely gluonic system. In this description,
our fundamental quantity is the gluon distribution function $f(\vb p,t)$, which is evolved in time using the Boltzmann equation
\begin{align}
\pdv[f(\vb p)]{\tau} =-\Conetwo[f(\vb p)]- \Ctwotwo[f(\vb p)] + \frac{p_z}{\tau} \pdv[f_{\vb p}]{p_z},\label{eq:boltzmann_equation}
\end{align}
where $\Ctwotwo$ describes elastic collisions, $\Conetwo$ summarizes inelastic interactions including the QCD analog of the LPM effect \cite{Arnold:2002ja}, 
and the last term accounts for the longitudinal expansion of the plasma along the beam direction \cite{Mueller:1999pi}. We take $z$ to be the beam axis and assume homogeneity in the $x-y$ plane as well as approximate boost invariance at midrapidity.
We consider $f(\vb p,t)$ to be the distribution function
averaged over the $\nu_g=2(\NC^2-1)$ polarization and color degrees of freedom.
Our numerical implementation builds on earlier works \cite{AbraaoYork:2014hbk, Kurkela:2015qoa, kurkela_2023_10409474}, where the distribution function is discretized using weight 
functions to conserve particle number and energy density for each sample exactly (see \app \ref{app:discretization}).

In this work, we use the same methods and formulas for the computation of $\Conetwo$ as previously \cite{AbraaoYork:2014hbk},
\begin{align}
    \begin{split}
        \Conetwo[f(\vb p)]&=\frac{(2\pi)^3}{2p^2\nu}\int_0^\infty\dd{p'}\dd{k'}\delta(p-p'-k')\gamma^p_{p',k'}\\
        &\quad\times\Big\{f(\vb p)(1+f(p'\vbphat))(1+f(k'\vbphat))\\
        &\qquad\quad-f(p'\vbphat)f(k'\vbphat)(1+f(\vb p))\Big\}\\
        &+\frac{(2\pi)^3}{p^2\nu}\int_0^\infty\dd{p'}\dd{k}\delta(p+k-p')\gamma^{p'}_{p,k}\\
        &\quad\times\Big\{f(\vb p)f(k\vbphat)(1+f(p'\vbphat))\\
        &\qquad-f(p'\vbphat)(1+f(k\vbphat))(1+f(\vb p))\Big\},
    \end{split}
\end{align}
with unit vectors $\vbphat = \vb p / p$.
The effective splitting/joining rates $\gamma^p_{p'k}$ interpolate between the Bethe-Heitler and LPM regime and are obtained via \cite{Arnold:2002zm}
\begin{subequations}
\begin{align}
    \gamma^{p}_{p',k}=\frac{p'^4+p^4+k^4}{p'^3p^3k^3}\mathcal F^{\vbphat}(p,p',k),
\end{align}
where $\mathcal F$ is given by
\begin{align}
    \mathcal F^{\vbphat}(p',p,k)=\frac{\lambda\nu}{8(2\pi)^3}\int\frac{\dd[2]{\vb h}}{(2\pi)^2}2\vb h\cdot \Re \vb F(\vb h),
\end{align}
with $\vb F(\vb h)$ the solution to the integral equation
\begin{align}
\begin{split}
    &2\vb h = i\delta E(\vb h; p',p,k)\vb F(\vb h)\\
    &+\frac{\lambda T_\ast}{2}\int\frac{\dd[2]{\vb \qperp}}{(2\pi)^2}\left(\frac{1}{\qperp^2}-\frac{1}{\qperp^2+m_D^2}\right)\\
    &\times\Big\{3 \vb F(\vb h)-\vb F(\vb h-k\vb \qperp)-F(\vb h-p'\vb \qperp)-F(\vb h-p\vb \qperp)\Big\}\label{eq:integral_equation}
    \end{split}
\end{align}
and 
\begin{align}
    \delta E(\vb h; p',p,k)=\frac{m_D^2}{4}\left(\frac{1}{k}+\frac{1}{p}+\frac{1}{p'}\right)+ \frac{\vb h^2}{2pkp'}.
\end{align}
\end{subequations}
The effective IR temperature $T_\ast$ is given by
\begin{align}
    T_\ast = \frac{8\lambda}{m_D^2}\int\frac{\dd[3]{\vb p}}{(2\pi)^3}f(\vb p)(1+f(\vb p)),\label{eq:Tstar}
\end{align}
and the Debye mass is defined as
\begin{align}
	m_D^2 = 8\lambda\int\frac{\dd[3]{\vb p}}{(2\pi)^32|\vb p|}f(\vb p),\label{eq:debye_mass_general}
\end{align}
with equilibrium value $m_D^2=g^2T^2\NC/3$.
Note that Eq.~\eqref{eq:integral_equation} involves an isotropic screening approximation, which allows to evaluate the longitudinal integral over the Wightman correlation function $\langle\langle A_\mu(K)A_\nu^\ast(K)\rangle\rangle$ using a sum rule \cite{Aurenche:2002pd}, as explained in more detail in Ref.~\cite{Arnold:2002zm}.
The inelastic collision kernel remains unchanged from previous implementations \cite{Kurkela:2014tea, Kurkela:2015qoa}.

Instead, in this paper, we study modifications of the elastic collision term $\Ctwotwo$, which is given by
\begin{align}
\Ctwotwo[f(\vb p)]&=\frac{1}{4|\vb{p}|\nu_g}\int_{\vb{kp'k'}}\left|\mathcal M(\vb{p},\vb{k};\vb{p'}\vb{k'})\right|^2 \nonumber\\
&\quad\times(2\pi)^4\delta^4(P+K-P'-K')\label{eq:c22_first}\\
&~~~\times\Big\{f(\vb p)f(\vb k)\left[1+ f(\vb p')\right]\left[1+ f(\vb k')\right]\nonumber\\
&~~\quad - f(\vb p')f(\vb k')\left[1+f(\vb p)\right]\left[1 + f(\vb k)\right]\Big\} \nonumber.
\end{align}
The Lorentz-invariant integration measure is defined as
\begin{align}
\int_{\vb k}:=\int\frac{\dd[3]{\vb k}}{(2\pi)^3 2k}.
\end{align}
At leading-order, the matrix element for $2\leftrightarrow 2$ gluon scattering reads
\begin{align}
\label{eq:gluonic_matrixel}
\frac{\left|\mathcal M\right|^2}{4 \lambda^2 \dA} =  9 + \frac{(t-s)^2}{\underline{u^2}} + \frac{(s-u)^2}{\underline{t^2}} + \frac{(u-t)^2}{s^2} \,.
\end{align}
It can be rewritten using the symmetry of exchanging $u$ and $t$ (corresponding to the exchange of the external particles with momenta $\vb p'$ and $\vb k'$) to
\begin{align}
\label{eq:gluonic_matrixel_simple}
\frac{\left|\mathcal M\right|^2}{4 \lambda^2 \dA} =  9 + 2{\frac{(s-u)^2}{\underline{t^2}}} + \frac{(u-t)^2}{s^2} \,.
\end{align}
Here, $s$, $t$, and $u$ are the usual Lorentz invariant Mandelstam variables, which are given in terms of the in- ($P$, $K$) and outgoing ($P'$, $K'$) momenta, $s=-(P+K)^2$, $t=-(P'-P)^2$, and $u=-(K'-P)^2$. Note that
we use the mostly plus metric convention here, $\eta_{\mu\nu}=\text{diag}(-1,1,1,1)$, and our results are, of course, independent of the specific choice.
We define the transferred momentum as $Q^\mu=(\omega, \vb q)=P'^\mu - P^\mu$.
As usual 
in EKT \cite{Arnold:2002zm}, we assume that all particle masses are negligible with respect to thermal masses $\sim gT$.
The Mandelstam variables then satisfy the usual relation
\begin{align}
    s+t+u=0.\label{eq:mandelstam_sum}
\end{align}
Additionally, also the momenta of the relevant excitations should be parametrically large compared to medium-dependent corrections,
which are included \cite{Arnold:2002zm} by replacing the underlined terms in Eq.~\eqref{eq:gluonic_matrixel} by%
\footnote{The replacement for the first underlined term in Eq.~\eqref{eq:gluonic_matrixel} is analogous with $u\leftrightarrow t$.}
\begin{align}
	{\frac{(s-u)^2}{\underline{t^2}}}&\to \left|\Gret_{\mu\nu}(P-P')\;(P+P')^\mu(K+K')^\nu\right|^2\label{eq:amy_replacement},
\end{align}
where $G_{\mu\nu}(Q)$ denotes the retarded propagator.
Note that when using the vacuum propagator $G_{\mu\nu}(Q)=g^{\mu\nu}/{Q^2}$ in \eq \eqref{eq:amy_replacement}, the right-hand side reduces to the left-hand side. We discuss the motivation for this replacement and its validity in more detail in the following.

\subsection{AMY-screening in the matrix element\label{sec:amy-screening-prescription}}
Since in this paper, our focus lies on different implementations to take into account medium effects, let us briefly review how the screening prescription
\eqref{eq:amy_replacement}
arises in Ref.~\cite{Arnold:2002zm}.
In a (thermal) medium, the propagators for the soft modes have to be resummed, whereas the hard modes can be treated as in vacuum. This is the basis for hard thermal loop (HTL) effective theory \cite{Braaten:1989mz}, and it has been applied to non-thermal media as well \cite{Mrowczynski:2004kv}. 

At tree level that is relevant for the leading-order matrix element in Eq.~\eqref{eq:gluonic_matrixel}, only one internal propagator $D$ appears.
The medium corrections to its free form $D_0$ are conventionally encoded in the self-energy,
\begin{align}
	\Pi_{\mu\nu}=(D^{-1})_{\mu\nu}-(D_0^{-1})_{\mu\nu}\,. \label{eq:self-energy}
\end{align}
In thermal equilibrium, this self-energy is proportional to the screening mass scale $m_D = \mathcal O(gT)$. 
Therefore, at leading order, medium effects are only relevant for soft momenta of $\mathcal O(gT)$.
In an anisotropic nonequilibrium system, one can still define an (isotropic) Debye mass $m_D$ via \eq \eqref{eq:debye_mass_general}.

Using Mandelstam variables, it is easy to see that internal soft momenta correspond to the case $|t|\ll s$. In \app \ref{app:kinematic-considerations} we show that this requirement implies that both $\omega, q \ll p, k$. Therefore, the small $t$ behavior is the region of interest for medium modifications.%
\footnote{We do not need to consider the small $u$ region here separately, since we have already used the symmetry $t \leftrightarrow u$ to rewrite the matrix element \eqref{eq:gluonic_matrixel_simple} such that only the small $t$ behavior needs to be regulated.}
\begin{align}
    0 < -t \ll s \approx -u.
\end{align}

At leading order,
the QCD $2\leftrightarrow2$ scattering matrix elements are spin-independent for soft momentum exchange (we refer to \app \ref{app:scattering_soft_gluon_exchange} for a detailed analysis). Therefore, medium effects can be included as in a theory with fictitious scalar `quarks', which can be regarded as a generalization of scalar QED.
For quark scattering in this fictitious scalar QCD,
the vertex factor is given by $-igt^a(P+K)^\nu$, where $P$ and $K$ are the momenta of the in- and outgoing quark, $g$ is the coupling constant and $t^a$ is a basis element of the $\mathfrak{su}(\NC)$ Lie algebra.
There are no spinor factors for external legs, and thus the matrix element (without color and coupling factors) is given by
\begin{align}
	\left|\mathcal M\right|^2\propto\left|(P+P')^\mu(K+K')^\nu G_{\mu\nu}(Q)\right]^2,\label{eq:scalar_quark_result}
\end{align}
which precisely corresponds to the screening prescription \eqref{eq:amy_replacement}. 
In \app \ref{app:scattering_soft_gluon_exchange}, we show explicitly that also for quark and gluon scattering with soft-gluon exchange, the simple screening form \eqref{eq:amy_replacement} is leading-order accurate.

\subsection{Isotropic HTL gluon propagator in a linear gauge\label{sec:isoHTLgeneral}}

While the HTL effective action itself is gauge-dependent, several physical quantities have been shown to be gauge-independent \cite{Kobes:1990dc, Kobes:1990xf}. In particular, in an isotropic system, the HTL gluon self-energy can always be written in terms of two functions $\Pi_a(Q)$ and $\Pi_b(Q)$.

In this section, we will discuss the most general form of the gluon propagator with HTL self-energy corrections in a linear gauge that does not break the rotational symmetry in the plasma rest frame. We will show that all gauge-dependent parts of the propagator are proportional to the exchange momentum $Q$ and will vanish when contracted with the external momenta in \eq \eqref{eq:amy_replacement}.
This confirms that the replacement \eqref{eq:amy_replacement} is gauge-independent.

First, let us denote the plasma rest frame by the vector $\tilde n^\mu=(1,\vb 0)$. The propagator depends on the 4-momentum $Q^\mu$. Thus, any quantity can be decomposed with respect to $\tilde n^\mu$, $Q^\mu$, and the metric $g^{\mu\nu}$. For simplicity, we will construct our tensor basis with the vector $n^\mu$ that is orthogonal to $Q^\mu$,
\begin{align}
        n_\mu &= P_{\mu\nu}\tilde n^\nu, & P_{\mu\nu}=g_{\mu\nu}-\frac{Q_\mu Q_\nu}{Q^2}.
\end{align}

In momentum space, we can represent any linear gauge condition \cite{Baier:1992mg, Bellac:2011kqa} $f^\mu A_\mu=0$
by a general vector $f^\mu(Q)$. If it does not break rotational invariance, we may decompose $f_\mu$ into a part parallel and transverse to $Q$,
\begin{align}
    f_\mu &= a(Q)Q_\mu + b(Q)n_\mu,
\end{align}
This general linear gauge includes the Lorenz (covariant) gauge, $f_\mu = Q_\mu$, and the Coulomb gauge, where $f_0 = 0$ and $f_i=q_i$, as well as the temporal axial gauge $f_0=\Lambda$ and $f_i=0$, where $\Lambda$ is a constant scale needed for dimensional reasons. Since the gluon propagator is symmetric and can only be built from the tensors $g_{\mu\nu}$, $Q_\mu$ and $n_\mu$, we may express it in the basis \cite{Bellac:2011kqa}
\begin{subequations}
\begin{align}
	  B_{\mu\nu}&=\frac{n_\mu n_\nu}{n^2}, &	C_{\mu\nu}&=n_\mu Q_\nu + n_\nu Q_\mu, \\
    E_{\mu\nu}&=\frac{Q_\mu Q_\nu}{Q^2}, &
	A_{\mu\nu}&=P_{\mu\nu}-B_{\mu\nu}.
\end{align}
\end{subequations}
The free propagator then reads \cite{Baier:1992mg, Bellac:2011kqa}
\begin{align}
    G^0_{\mu\nu}=\frac{P_{\mu\nu}}{Q^2}-\frac{1}{f_e Q^2}\left(f_c C_{\mu\nu}+(f_b+ Q^2)E_{\mu\nu}\right),
\end{align}
with
\begin{align}
    f_b=\frac{b^2n^2}{\xigauge},&& f_c=\frac{ab}{\xigauge}, && f_e=\frac{a^2Q^2}{\xigauge},\label{eq:f_relations}
\end{align}
where $\xigauge$ is the gauge-fixing parameter.
Including the self-energy \eqref{eq:self-energy} leads to the dressed propagator,
\begin{align}
    G_{\mu\nu}&=\frac{A_{\mu\nu}}{Q^2+\Pi_a}+\frac{B_{\mu\nu}}{\tilde b(Q)}-\frac{f_c+\Pi_c}{\tilde b(Q)(f_e+\Pi_e)}C_{\mu\nu}\nonumber\\
    &\qquad+\frac{Q^2+\Pi_b+f_b}{\tilde b(Q)(f_e+\Pi_e)}E_{\mu\nu},
\end{align}
with
\begin{align}
    \tilde b(Q)&=Q^2+\Pi_b+f_b-n^2Q^2\frac{(f_c+\Pi_c)^2}{f_e+\Pi_e},
\end{align}
and where we have also decomposed the self-energy 
\begin{align}
    \Pi_{\mu\nu} = \Pi_a A_{\mu\nu} + \Pi_b B_{\mu\nu} + \Pi_c C_{\mu\nu} + \Pi_e E_{\mu\nu}.
\end{align}
The Ward identity forces the HTL self-energy to be transverse, $Q^\mu \Pi_{\mu\nu}=0$. In that case, one finds that $\Pi_c=\Pi_e=0$, which simplifies $\tilde b(Q)=Q^2+\Pi_b$. All dependence from the gauge choice is now in the parameters $f_b$, $f_c$, and $f_e$, which always appear together with factors of $Q_\mu$.
A closer inspection reveals that all terms proportional to $Q_\mu$ or $Q_\nu$ yield zero when contracted with the external momenta in Eq.~\eqref{eq:amy_replacement}. This follows from
\begin{align}
    Q \cdot (P+P')&=(P'-P)\cdot(P+P')=P'^2-P^2=0,
\end{align}
and similarly with $P\leftrightarrow K$.
Therefore, the screening prescription in Eq.~\eqref{eq:amy_replacement} is gauge invariant.

\subsection{Expectation values and observables}
The expectation value of an observable $O(\vb p,t)$ for a given distribution function $f(\vb p, t)$ can be obtained via
\begin{align}
    \langle O(t)\rangle = \frac{1}{n(t)}\int\frac{\dd[3]{\vb p}}{(2\pi)^3}\, O(\vb p, t) f(\vb p, t),
\end{align}
with the particle number density
\begin{align}
n(t)=\int\frac{\dd[3]{\vb p}}{(2\pi)^3}f(\vb p,t).
\end{align}

An important class of such observables 
are the components of the energy-momentum tensor,
\begin{align}
    T^{\mu\nu}=2(\NC^2-1)\int\frac{\dd[3]{\vb p}}{(2\pi)^3} \frac{p^\mu p^\nu}{|\vb p|}f(\vb p),
\end{align}
whose diagonal entries correspond to the energy density $\varepsilon = T^{00}=\langle p\rangle$, transverse $P_T=T^{xx}=T^{yy}$ and longitudinal pressure $P_L = T^{zz}$.
Since we consider a conformal theory, we have $\varepsilon = 2P_T+P_L$, and in an isotropic system $P_L=P_T=\varepsilon/3$.

\section{Screening prescriptions\label{sec:different_screening_prescriptions}}

In the discussion of the Boltzmann equation in \se \ref{sec:Boltzmann} we have mentioned that we need to regulate unphysical divergences at low momenta of the matrix element $|\mathcal M|^2$, and especially its term $(s-u)^2/\underline{t^2}\,$. Here we will introduce the screening prescriptions for \eq \eqref{eq:amy_replacement} that we will study in this work.

\subsection{Debye-like screening\label{sec:debye-like-screening}}

The usually implemented prescription uses 
a simple Debye-like screened propagator \cite{AbraaoYork:2014hbk},
\begin{align}
    G_{\mu\nu}=\frac{g_{\mu\nu}}{Q^2} \,S(q),
    \label{eq:simple-isotropic-screening}
\end{align}
which at the level of the matrix elements  corresponds to 
\begin{align}
	\frac{(s-u)^2}{t^2}\to \frac{(s-u)^2}{t^2} \, S^2(q).
    \label{eq:simple_replacement}
\end{align}
The screening is implemented in the function
\begin{align}
    \label{eq:screening}
    S(q) = \frac{q^2}{q^2+\xiscreensqr \, m_D^2},
\end{align}
where the Debye mass is given by Eq.~\eqref{eq:debye_mass_general}.
The constant $\xiscreen=e^{5/6}/\sqrt{8}$ is chosen to approximate the HTL propagator 
for longitudinal momentum broadening 
or in isotropic systems \cite{AbraaoYork:2014hbk, Boguslavski:2023waw}.
This simple Debye-like screening prescription is typically used in EKT simulations \cite{Kurkela:2015qoa, Kurkela:2018xxd, Du:2020zqg}.

As explained in \se\ref{sec:amy-screening-prescription}, screening effects need only be included for $|t|\ll s \approx -u$. Therefore, also other screening prescriptions are possible.
Here, we show that multiplying this screening factor $S(q)$
with another term as $su/t^2$ or $s^2/t^2$ is also leading-order accurate.

To be more precise, one can easily show that the three expressions
\begin{align}
	I_1=\frac{(s-u)^2}{4t^2}, && I_2 = -\frac{su}{t^2}, && I_3=\frac{s^2}{t^2}.\label{eq:possible-screening-mandelstam-bare}
\end{align}
are equivalent up to $\mathcal O(|t|/s)$, in particular, $I_3/I_1=1+\mathcal \mathcal O(|t|/s)$ and $I_2/I_1=1+\mathcal O(t^2/s^2)$. 

We may now implement an analogous screening prescription to \eqref{eq:simple_replacement} in any of the terms \eqref{eq:possible-screening-mandelstam-bare} entering the gluonic matrix element in the formulations
\begin{subequations}\label{eq:equivalent-matrixelements}
\begin{align}
 \frac{|\mathcal M|^2}{16d_AC_A^2g^4}
 &=\frac{1}{4}\left(9+2\frac{(s-u)^2}{t^2}+\frac{(t-u)^2}{s^2}\right)\\
 &=3-2\frac{su}{t^2}-\frac{tu}{s^2}\\
 &=3+2\frac{s^2}{t^2}+2\frac{s}{t}-\frac{tu}{s^2}\,.
\end{align}
\end{subequations}
Since all of these vacuum matrix elements \eqref{eq:equivalent-matrixelements} are equivalent due to \eqref{eq:mandelstam_sum}, and to leading-order (or up to $\mathcal O(|t|/s)$) the expressions in \eqref{eq:possible-screening-mandelstam-bare} coincide, we have a particular freedom of how to implement the screening. The choice mentioned in Ref.~\cite{Arnold:2002zm} is to implement the screening as in the scalar quark case, Eq.~\eqref{eq:scalar_quark_result}, which makes the gauge-independence manifest.

In this paper, we study the inclusion of the simple isotropic screening in Eq.~\eqref{eq:screening} to the different equivalent expressions \eqref{eq:possible-screening-mandelstam-bare} in \eqref{eq:equivalent-matrixelements}, by using the matrix elements,
\begin{subequations}\label{eq:different_regularization_matrix_element}\begin{align}
	\frac{|\Mdebyeone|^2}{16d_AC_A^2g^4}&=\frac{1}{4}\left(9+2\frac{(s-u)^2}{t^2} \, S^2(q)+\frac{(t-u)^2}{s^2}\right)\label{eq:usual_screened_matrix_element}\\
	\frac{|\Mdebyetwo|^2}{16d_AC_A^2g^4}&=3-2\frac{su}{t^2}\, S^2(q)-\frac{tu}{s^2}
 \label{eq:Debye2_screened_matrix_element}\\
	\frac{|\Mdebyethree|^2}{16d_AC_A^2g^4}&=3+2\left(\frac{s^2}{t^2}+\frac{s}{t}\right) S^2(q)-\frac{tu}{s^2}\,.
 \label{eq:Debye3_screened_matrix_element}
\end{align}
\end{subequations}
Because of Eq.~\eqref{eq:mandelstam_sum}, the second and third matrix elements coincide, $\Mdebyetwo\equiv\Mdebyethree$, and we will only consider differences between $\Mdebyeone\neq \Mdebyetwo$.

Note that for the region $|t|\ll s$, 
the agreement of the matrix elements \eqref{eq:different_regularization_matrix_element}
is independent of the values of $q$ and $m_D$. On the other hand, for sufficiently large $q \gg m_D$ and independent of the values of $s,t$, these matrix elements only differ up to factors of $\mathcal O\left(m_D^2/q^2\right)$.
Since screening effects are only important for these regions of small $q\sim m_D\sim gT$ or small $|t|\ll s$, the matrix elements \eqref{eq:different_regularization_matrix_element} are leading-order equivalent.

\subsection{isoHTL screening}
We also include the full isotropic HTL retarded propagator in Eq.~\eqref{eq:amy_replacement},
\begin{align}
    &\frac{|\Mhtl|^2}{16d_AC_A^2g^4}=\frac{1}{4}\bigg(9+\frac{(t-u)^2}{s^2}\label{eq:full_isotropic_HTL_matrixelement}\\
    &\qquad +2\left|\GretHTL_{\mu\nu}(P-P')(P+P')^\mu(K+K')^\nu\right|^2\bigg).\nonumber
\end{align}
We will refer to this screening prescription as isoHTL. As we have discussed in \se\ref{sec:isoHTLgeneral}, Eq.~\eqref{eq:full_isotropic_HTL_matrixelement} is gauge invariant. For convenience, we choose to use the isotropic HTL propagator in strict Coulomb gauge,
\begin{subequations}
\begin{align}
    \GLR(Q)&=\frac{1}{q^2+\PiLR(\omega/q)},\\
    \GTR(Q)&=\frac{-1}{q^2-\omega^2+\PiTR(\omega/q)},
\end{align}
\end{subequations}
and $G^{ij}(Q)=\left(\delta^{ij}-\frac{q^iq^j}{q^2}\right)\GTR(Q)$, with the self-energies given by
\begin{subequations}
\begin{align}
    \RE\PiLR(x)&=m_D^2\left(1-\frac{x}{2}\ln\left|\frac{x+1}{x-1}\right|\right),\\
    \IM\PiLR(x)&=\frac{xm_D^2\pi}{2}\theta(1-|x|),\\
    \RE \PiTR(x)&=\frac{m_D^2}{2}-\frac{1}{2}(1-x^2)\RE\PiLR(x),\\
    \IM\PiTR(x)&=-\frac{1}{2}(1-x^2)\IM\PiLR(x),
\end{align}
\end{subequations}
where $x=\omega/q$. Kinematically, $|x|=|\omega|/q < 1$, and thus the imaginary part is always nonzero for $\omega\neq 0$. Furthermore, $G(-Q)$ corresponds to the advanced propagator, and $\IM\Pi(-Q)=-\IM\Pi(Q)$.
Contracting the propagator with the external momenta yields
\begin{align} 	
&\left|\GretHTL_{\mu\nu}(P-P')(P+P')^\mu (K+K')^\nu\right|^2 \label{eq:HTL_propagator_explicit_expression}
\\
&\qquad=\frac{c_1^2}{A^2+B^2}+\frac{c_2^2}{C^2+D^2}-\frac{2c_1c_2(AC+BD)}{(A^2+B^2)(C^2+D^2)},\nonumber
\end{align}
with
\begin{subequations}
\begin{align}
    c_1&=(2p+\omega)(2k-\omega)\\
    c_2&=4pk\sin\theta_{qp}\sin\theta_{qk}\cos(\phi_{qk}-\phi_{qp}),\label{eq:constant_c2}\\
    A &= q^2+\RE\PiLR(x), \qquad\qquad B=\IM\PiLR(-x),\\
    C &= q^2-\omega^2+\RE\PiTR(x), \qquad D =\IM\PiTR(-x).
\end{align}
\end{subequations}
Here, $\thetaqp$ and $\thetaqk$ denote the angles between $\vb q$ and $\vb p$ or $\vb k$. Additionally, the angle $\phiqk$ ($\phiqp$) is the azimuthal angle of $\vb k$ ($\vb p)$ in a coordinate system where $\vb q$ points in the $z$ direction and the beam axis lies in the $xz$ plane. More details on the kinematic variables are summarized in \app\ref{app:kinematic-considerations}.%
\footnote{
Note that in \re \cite{Boguslavski:2023waw} the argument in the cosine in \eqref{eq:constant_c2} slightly differs due to a different frame choice.
}

\subsection{Debye-like screening as an approximation to isoHTL}
\label{sec:debye_as_approx}

The Debye-like screening prescription \eqref{eq:simple_replacement} can be understood as a simple approximation to the full isotropic HTL matrix element \eqref{eq:HTL_propagator_explicit_expression}. In the original work \cite{AbraaoYork:2014hbk} its screening parameter
$\xiscreen$ was chosen to reproduce (longitudinal) soft momentum transfer in elastic collisions. This can be seen as follows:
For isotropic distributions, the collision kernel can be written as%
\footnote{This form can be obtained analogously to the jet quenching parameter $\qhat$, as presented in Ref.~\cite{Boguslavski:2023waw}.} \cite{AbraaoYork:2014hbk}
\begin{align}
\begin{split}
    \Ctwotwo&=\frac{1}{2^9\pi^5\nu}\int_0^\infty\dd{k}\int_0^{2\pi}\dd{\phiqp}\\
    &\times\int_{-p}^k\dd{\omega}\Bigg\{f(p)f(k)(1+f(p+\omega))(1+f(k-\omega)) \\
    &\qquad - f(p+\omega)f(k-\omega)(1+f(p))(1+f(k))\Bigg\}\\
    &\times\int_{|\omega|}^{\min(2k-\omega,2p+\omega)}\dd{q}\int_0^{2\pi}\dd{\phiqk}\frac{\left|\mathcal M\right|^2}{p^2}.
    \end{split}
\end{align}
Screening effects are only important for soft internal momenta, $q,\,\omega\ll k,\,p$. When expanding the distribution functions for small $\omega$, the first nonvanishing term is quadratic
in $\omega$ since the matrix element is even. One therefore requires \cite{AbraaoYork:2014hbk} that in this limit
\begin{align}
    \int_{-\infty}^\infty\dd{\omega}\omega^2\int_{|\omega|}^\infty\dd{q}\int_0^{2\pi}\dd\phi \left(\left|\Mhtl\right|^2-\left|\Mdebyeone\right|^2\right)=0,
\end{align}
which fixes the constant $\xiscreen=e^{5/6}/\sqrt{8}$.

Moreover, in \re \cite{Boguslavski:2023waw} it was shown that the Debye-like screening prescription cannot simultaneously approximate both longitudinal and transverse momentum diffusion in isotropic systems, while the isoHTL prescription can. 
Therefore, the isoHTL screening prescription is more general and should be used when both processes are important.

Note, however, that neither of these screening prescriptions correctly includes the effect of plasma instabilities, which are generically seen to occur in anisotropic systems \cite{Mrowczynski:1988dz, Arnold:2003rq, Romatschke:2003ms, Romatschke:2006bb, Kurkela:2011ub, Hauksson:2021okc}. However, numerical evidence indicates that these instabilities do not play a dominant role at the time scales of interest for kinetic theory simulations \cite{Berges:2013eia, Berges:2013fga} and when a quasiparticle picture has become applicable \cite{Boguslavski:2018beu, Boguslavski:2021buh, Boguslavski:2021kdd}. 

\subsection{Extrapolating EKT to large couplings\label{sec:ekt_extrapolation_large_couplings}}

Although QCD kinetic theory is only valid at weak couplings, it is often extrapolated to larger values of the coupling in lieu of a fully dynamical quantum field theory simulation. Since the Debye mass $m_D$ is proportional to the coupling, the na\"ive extrapolation $\lambda\to\infty$ leads to $m_D\to\infty$, and thus the screened $t$-channel small-angle-scattering terms in the matrix element \eqref{eq:usual_screened_matrix_element} vanish and only the $s$-channel and constant parts remain.
Therefore, at sufficiently large couplings, differences in the screening prescription should become less important.
Although kinetic theory and perturbative QCD are not valid at large couplings, Eq.~\eqref{eq:boltzmann_equation} is still mathematically well-defined, often employed as a model of initial stages, and its numerical solution may provide phenomenological insights into the equilibration process in 
heavy-ion collisions.

\section{Thermalization of isotropic systems\label{sec:thermalization_isotropic}}

\begin{figure*}
    \centerline{
    \includegraphics[width=0.47\linewidth]{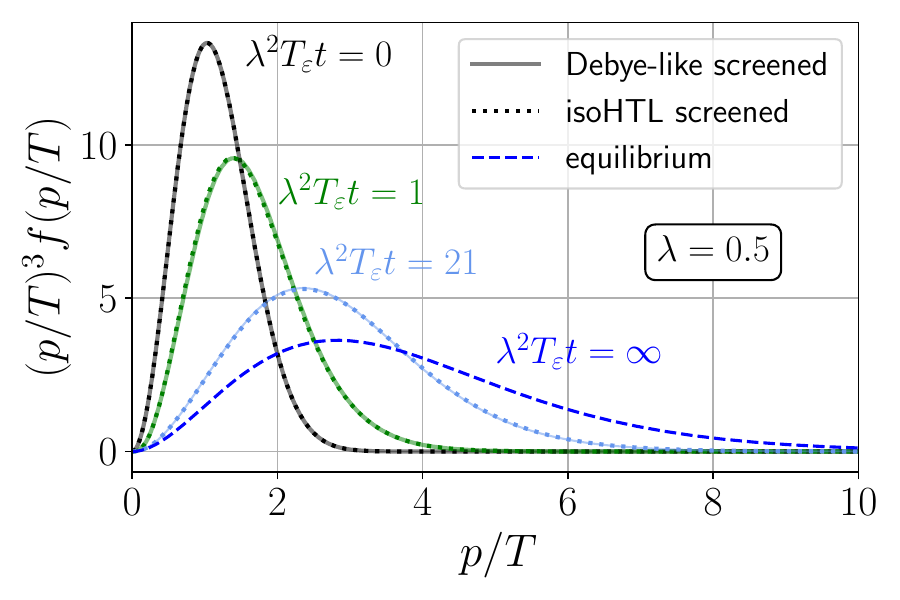}
    \includegraphics[width=0.47\linewidth]{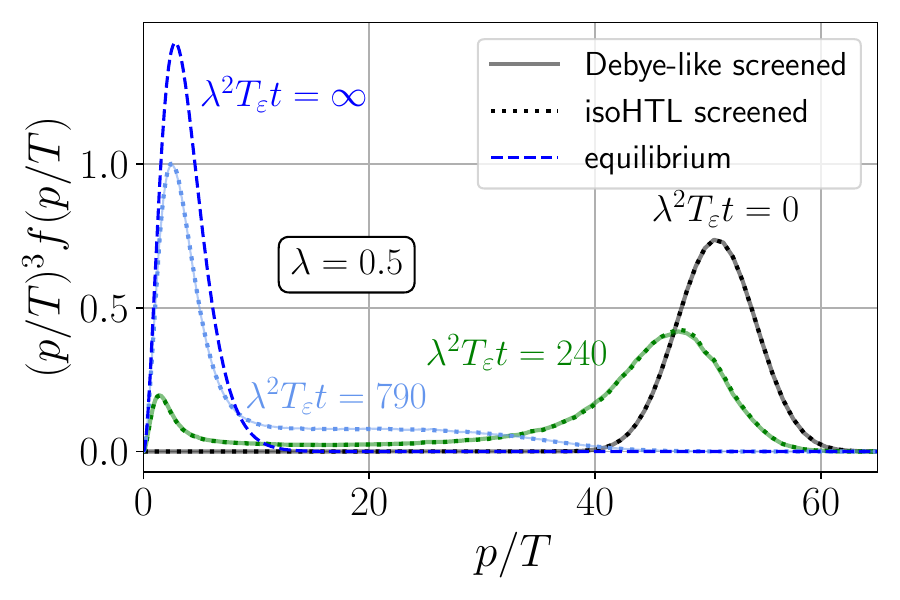}
    }
    \centerline{
    \includegraphics[width=0.47\linewidth]{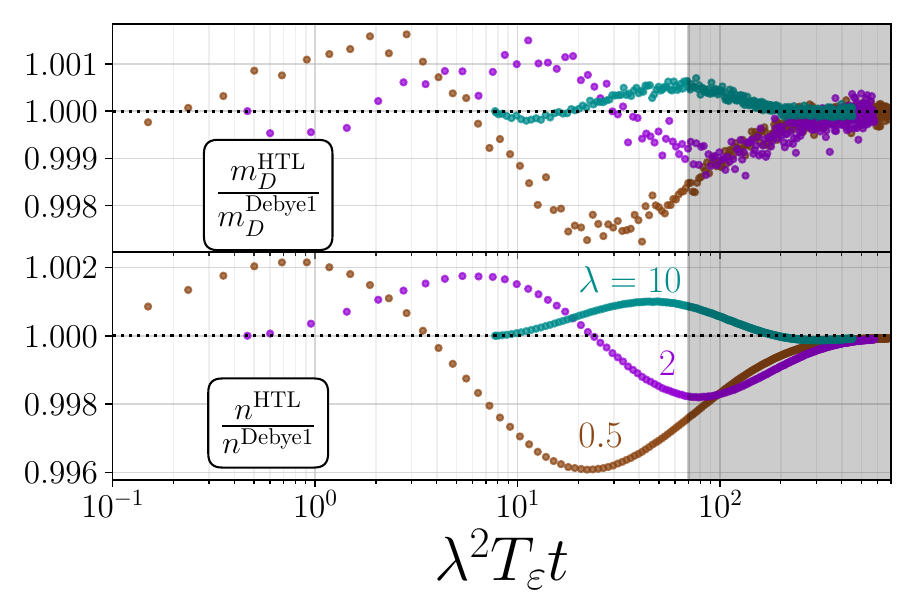}
    \includegraphics[width=0.47\linewidth]{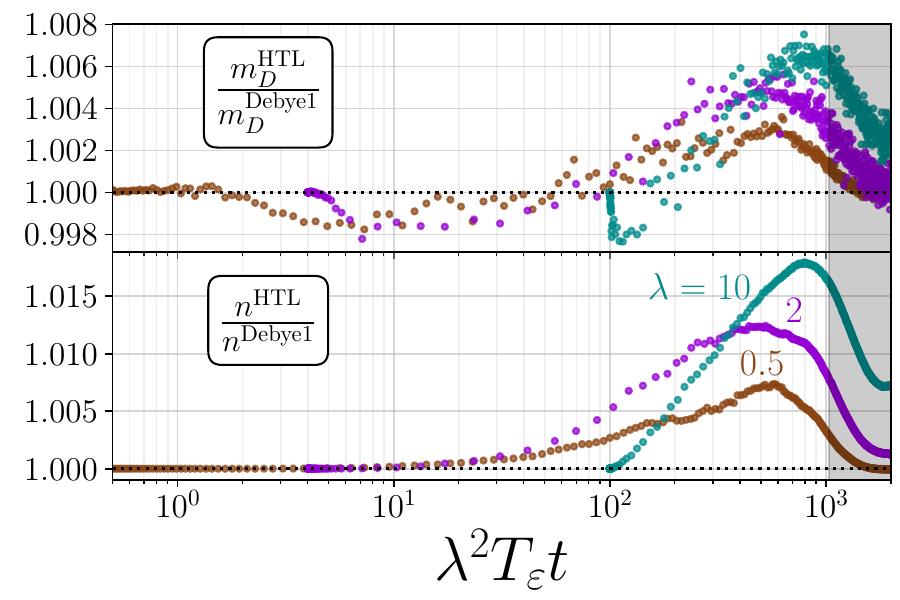}
    }
    
    \caption{Thermalization of an overoccupied (left) and underoccupied (right) gluonic system. In the top row, we show a scaled distribution function $p^3f$ as a function of momentum whereas in the bottom row, we show ratios of observables (Debye mass $m_D$ and particle number $n$) calculated with isoHTL screening \eqref{eq:full_isotropic_HTL_matrixelement} over Debye-like 1 screening \eqref{eq:usual_screened_matrix_element} as functions of scaled time $\lambda^2 \Teps t$. 
    }
    \label{fig:iso_thermalization}
\end{figure*}

\begin{table}
\begin{ruledtabular}
\begin{tabular}{c c c}
$\lambda$ & Debye-like: $\lambda^2 T \ttherm$ & isoHTL: $\lambda^2T \ttherm$\\ [0.5ex]
\hline
$0.5$ & $69.7$ & $67.0$\\
$2$ & $86.0$ & $84.4$\\
$10$ & $93$ & $93$
\\[1ex]
\end{tabular}
\end{ruledtabular}
\caption{Thermalization times for overoccupied systems (corresponding to the left column of \fig \ref{fig:iso_thermalization}).
}
\label{tab:UV_thermalizationtimes}
\end{table}
\begin{table}
\begin{ruledtabular}
\begin{tabular}{c c c}
$\lambda$ & Debye-like: $\lambda^2 T \ttherm$ & isoHTL: $\lambda^2T \ttherm$ \\ [0.5ex]
\hline
$0.5$ & $1029$ & $1022$\\
$2$ & $1127$ & $1112$\\
$10$ & $1245$ & $1216$
\\[1ex]
\end{tabular}
\end{ruledtabular}
\caption{Thermalization times for underoccupied systems (corresponding to the right column of \fig \ref{fig:iso_thermalization}).
}
\label{tab:IR_thermalizationtimes}
\end{table}

The approach to equilibrium of an initially far-from-equilibrium system of gluons has been studied extensively using EKT \cite{Kurkela:2014tea, Fu:2021jhl} and has also been extended to include quarks \cite{Kurkela:2018oqw, Kurkela:2018xxd} and non-zero baryon chemical potential \cite{Du:2020dvp, Du:2020zqg}.
In this section, we re-evaluate the equilibration process of non-expanding isotropic systems where the longitudinal expansion term in \eq \eqref{eq:boltzmann_equation} is omitted by comparing results with our different screening prescriptions.%
\footnote{Since an isotropic distribution does not suffer from plasma instabilities that complicate the theory in anisotropic plasmas, our results with the HTL self-energy relaxes the Debye-like screening approximation originally employed and is leading-order correct.}
Similarly to \cite{Kurkela:2014tea, Fu:2021jhl}, we start with both an initially under- and overoccupied state. For the underoccupied system, we use the initial condition 
\begin{align}
    f(p)=A\exp\left(-(p-Q)^2/(Q/10)^2\right),
\end{align}
with $A$ chosen such that $Q=50 T$, where $T$ is the temperature of the equilibrium system after thermalization.
As an initial condition for the overoccupied system, we choose a parametrization of the system's self-similar scaling solution \cite{AbraaoYork:2014hbk}
\begin{align}
    f(p)=\frac{(Qt)^{-4/7}}{\lambda \tilde p}\left(0.22 e^{-13.3\tilde p}+2 e^{-0.92 \tilde p^2}\right),
\end{align}
with $\tilde p=(p/Q)(Qt)^{-1/7}$ and initialize our system at $Qt=0.12$.

\begin{figure*}
    \centering
    \centerline{ %m, lambdaTs, n
        \includegraphics[width=0.33\linewidth]{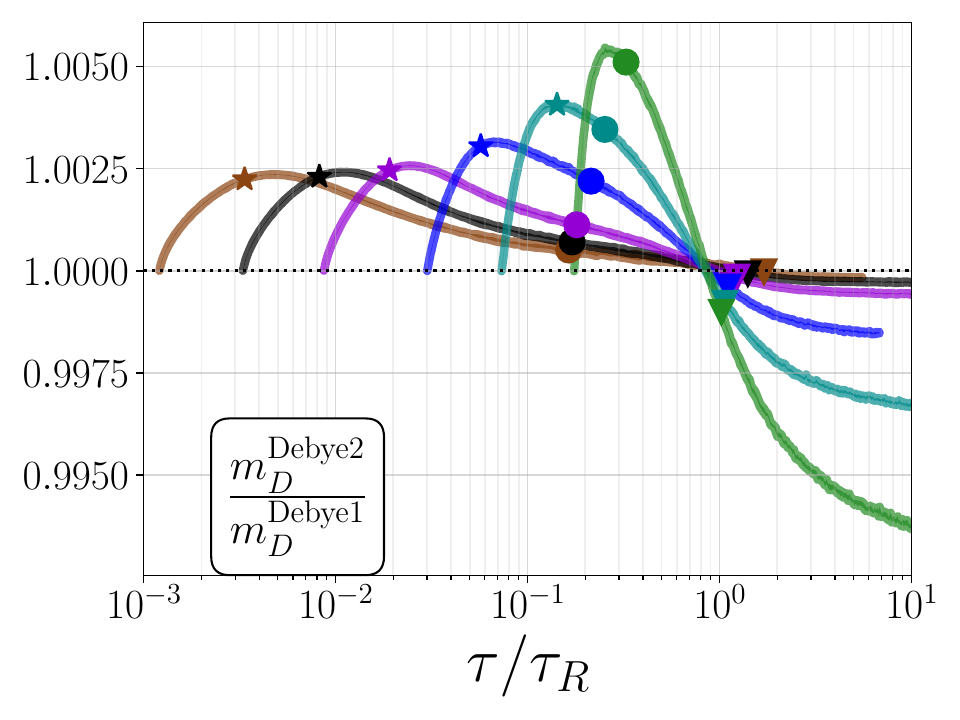}
        \includegraphics[width=0.33\linewidth]{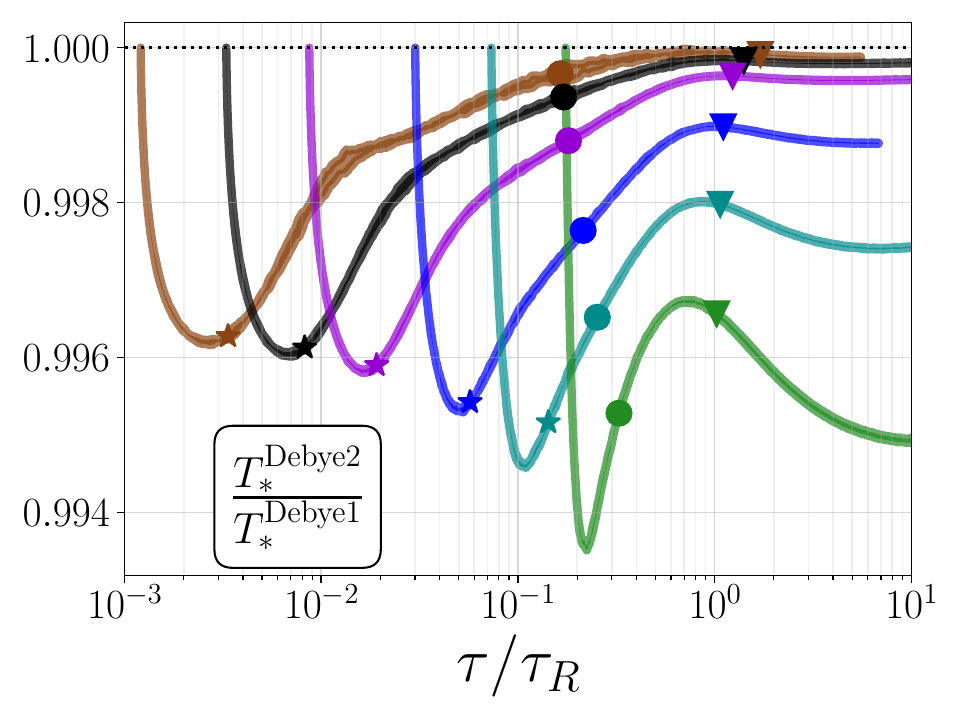}
        \includegraphics[width=0.33\linewidth]{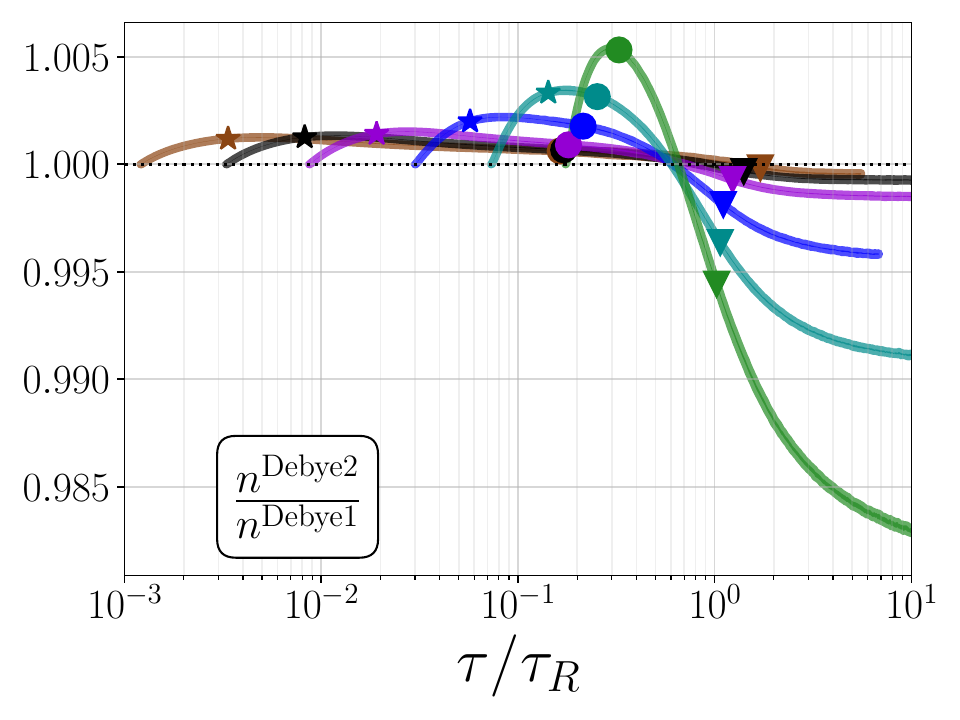}
    }
    \centerline{ % e, PT, PZ
        \includegraphics[width=0.33\linewidth]{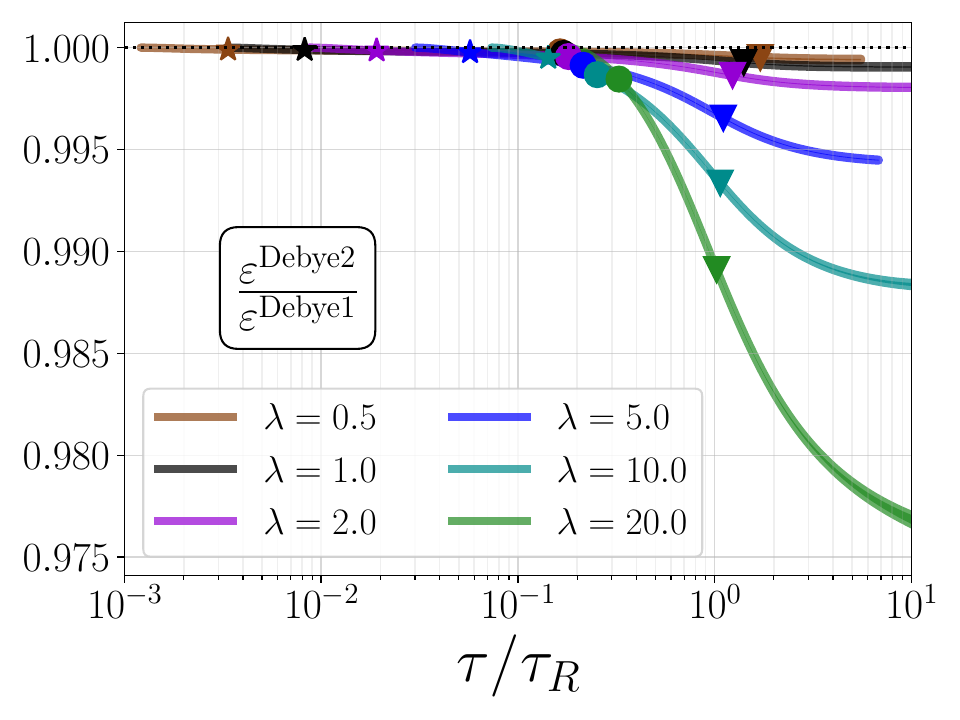}
        \includegraphics[width=0.33\linewidth]{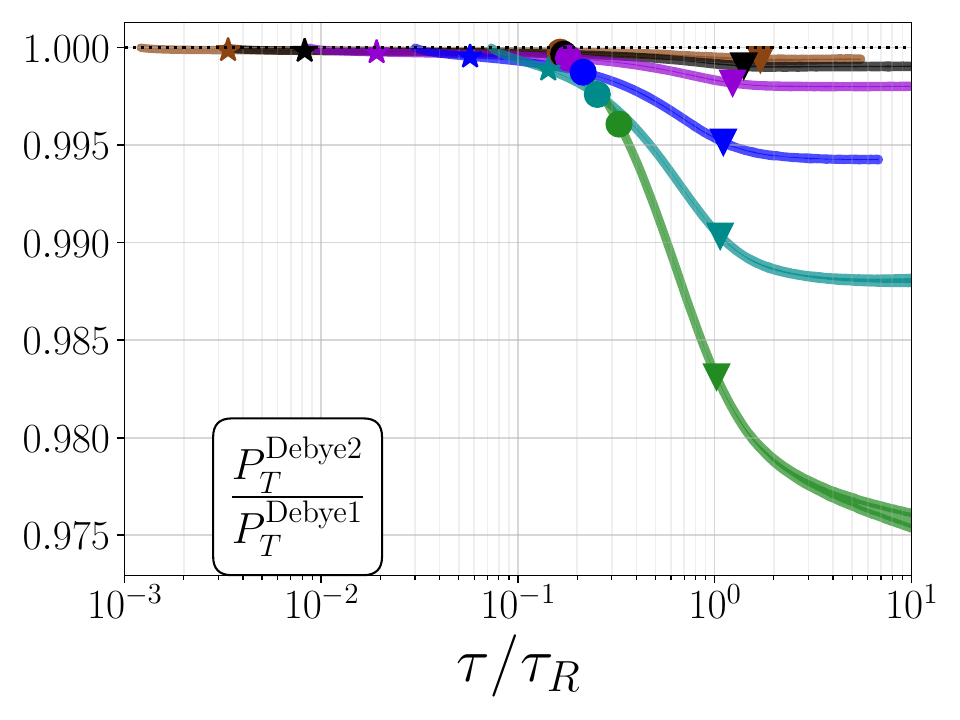}
        \includegraphics[width=0.33\linewidth]{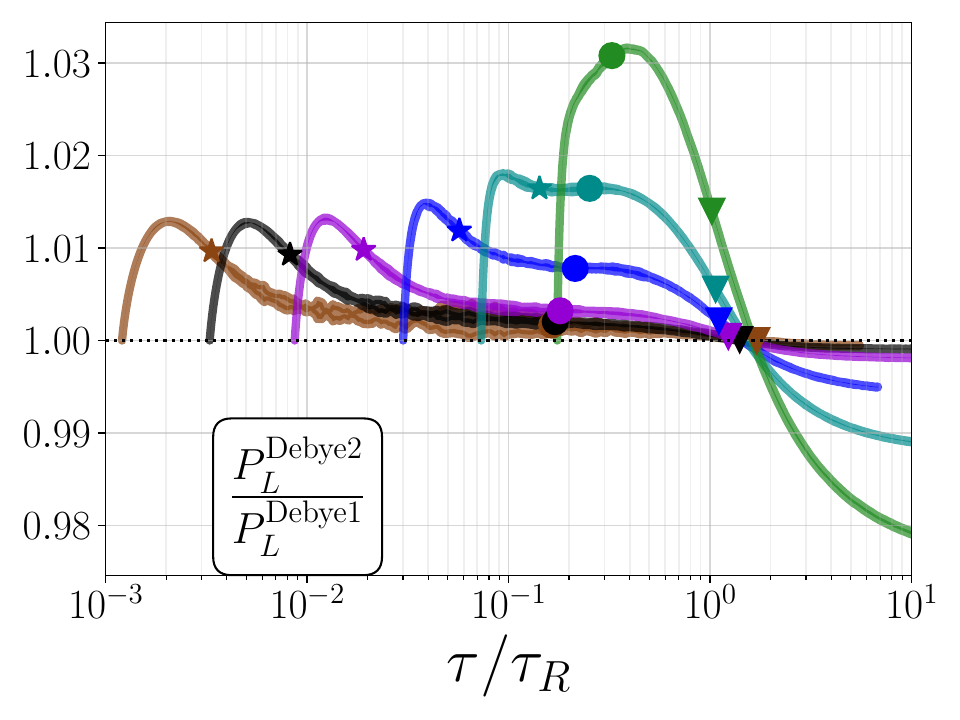}
    }
    \centerline{ % f, pt2, pz2
        \includegraphics[width=0.33\linewidth]{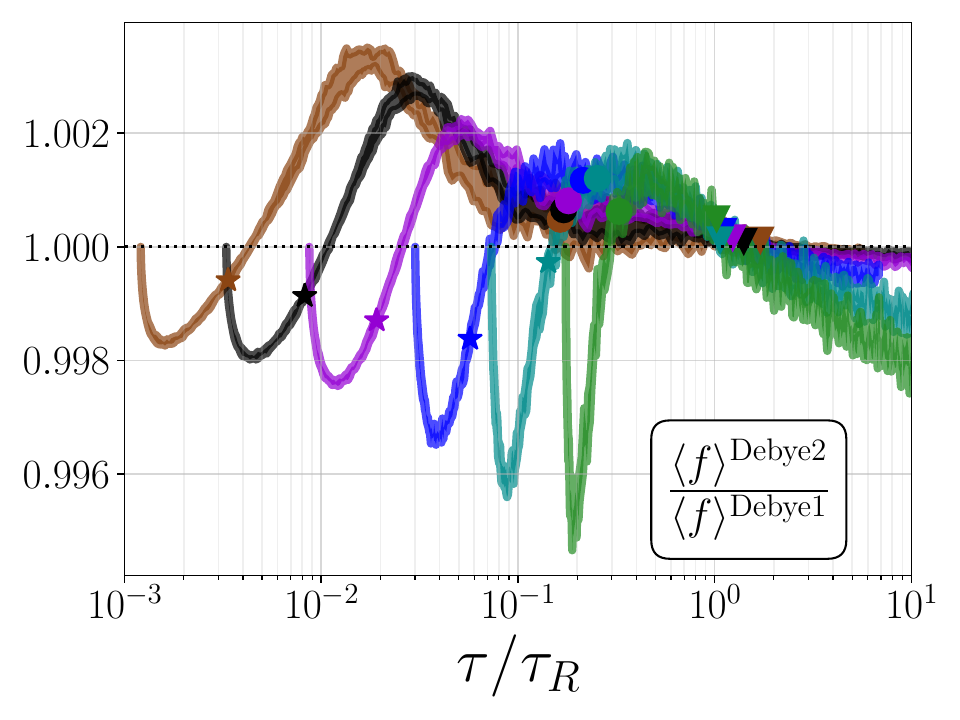}
        \includegraphics[width=0.33\linewidth]{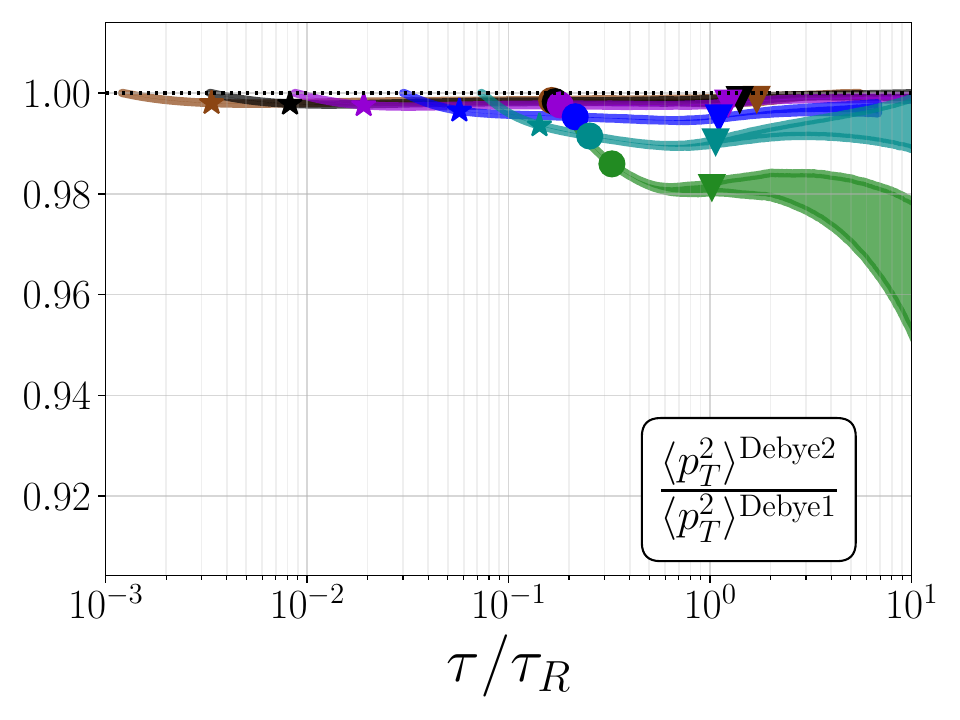}
        \includegraphics[width=0.33\linewidth]{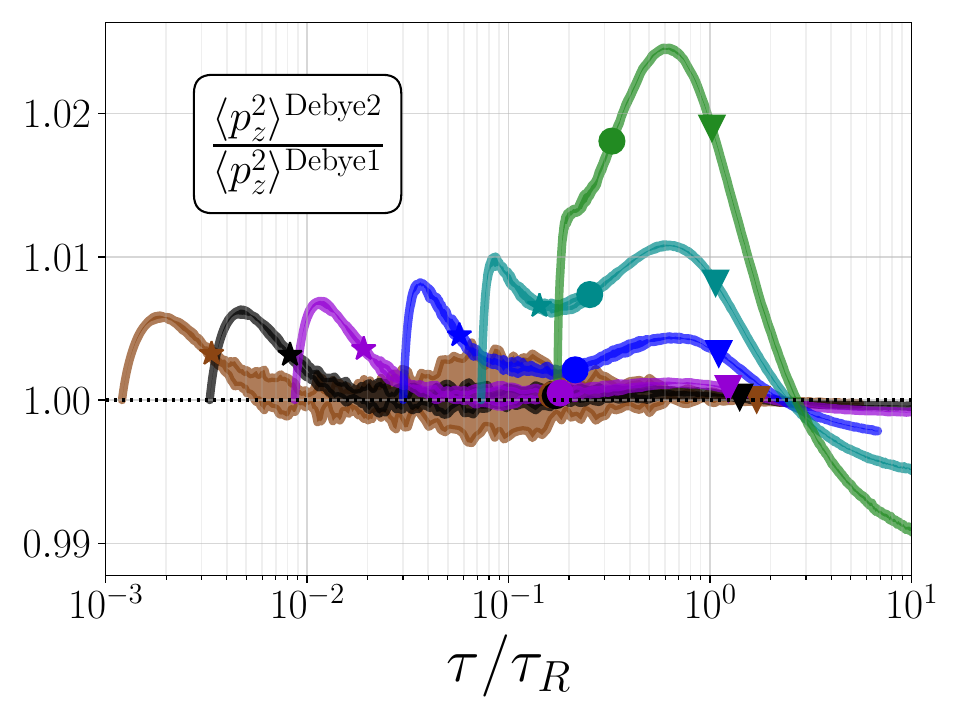}
    }
    \caption{Ratios of various observables when comparing simulations with the Debye-like screening 2 in comparison to Debye-like screening 1 in \eq \eqref{eq:different_regularization_matrix_element}. 
    The bands denote estimates for the standard deviation, obtained from a statistical average over five independent simulations.
    }
    \label{fig:debye-like-comparison}
\end{figure*}

By solving the Boltzmann equation \eqref{eq:boltzmann_equation} numerically, we follow the system's time evolution until equilibrium is reached, using a Debye-like $\Mdebyeone$ and isoHTL screened matrix element $\Mhtl$.
The system's time evolution can be fully described by the one-dimensional distribution function, whose rescaled version $p^3 f(p,t)$ is depicted in the top row of \fig\ref{fig:iso_thermalization} for over- (left) and underoccupied systems (right).
We find that both Debye-like (full lines) and isoHTL screening (dotted lines) lead to almost identical results since the curves nearly coincide. 

Next, we quantify the deviations between the screening prescriptions. We observe only a slight decrease in the thermalization time, which can be defined implicitly in terms of effective temperatures by \cite{Fu:2021jhl}
\begin{align}\left(T_0(\ttherm)/T_1(\ttherm)\right)^{\pm 4}=0.9,\label{eq:def_thermalizationtime}
\end{align}
with $+$ for under- and $-$ for overoccupied systems. The effective temperatures are defined via
\begin{align}
    T_\alpha=\left[\frac{2\pi^2}{\Gamma(\alpha+3)\zeta(\alpha+3)}\int\frac{\dd[3]{\vb p}}{(2\pi)^3}p^\alpha f(p)\right]^{1/(\alpha+3)}.
\end{align}
The first moment $T_1=\Teps$ has the physical interpretation of 
the temperature of a thermal system with the same energy density (Landau matching),
\begin{align}\Teps(\tau)=\left(\frac{30\varepsilon(\tau)}{\pi^2\nu}\right)^{1/4}.\label{eq:Teps}
\end{align}
Due to energy conservation in non-expanding systems, it is constant throughout the evolution and corresponds to the temperature of the thermal system after equilibration. The other effective temperature, $T_0$, is related to the particle number density $n$, via the relation 
$T_0=\left(\pi^2 n/\zeta(3)\right)^{1/3}$.
The thermalization times are listed in \tab \ref{tab:UV_thermalizationtimes} and \ref{tab:IR_thermalizationtimes}. We find that the full HTL matrix element leads to only slightly smaller thermalization times. 

We can also quantify differences in the dynamics. In the bottom row of \fig\ref{fig:iso_thermalization}, we show the evolution of the Debye mass \eqref{eq:debye_mass_general} and number density $n$ as ratios of a simulation with isoHTL screening over one with Debye-like-1 screening. 
The gray area indicates when the system is close to equilibrium, i.e., after the thermalization time as defined in \eqref{eq:def_thermalizationtime}.
We observe that both the Debye mass and the number density differ at a sub-percent level between the simulations.

We note that the very good agreement for isotropic systems among the screening prescriptions is not surprising.
As detailed in \se \ref{sec:debye_as_approx}, Debye-like screening was introduced specifically in the isotropic case to approximate (isotropic) HTL screening.

\section{Results\label{sec:results} with longitudinal expansion}

We now turn to systems undergoing Bjorken expansion as described by the Boltzmann equation \eqref{eq:boltzmann_equation}, which is relevant for the dynamical description of the plasma in heavy-ion collisions.

\subsection{Initial conditions, time markers and observables}

We use the same initial conditions as in Ref.~\cite{Kurkela:2015qoa},
\begin{align}  
    f(p_\perp,p_z,\tau{=}1/Q_s)=\frac{2A(\xianiso)\langle p_T\rangle}{\lambda \, p_\xi}\exp\left({-\frac{2p_\xi^2}{3\langle p_T\rangle^2}}\right), 
    \label{eq:initial_cond}
\end{align}
with $p_\xi=\sqrt{p_\perp^2+(\xianiso p_z)^2}$, $\xianiso = 10$, $A(\xi_0) =  5.24171$ and $\langle p_T\rangle = 1.8 Q_s$ as in previous studies.
Note that for consistency with the literature, we use the letter $\xianiso$ to denote the anisotropy parameter of the initial condition. It should not be confused with the gauge fixing parameter $\xigauge$ or the Debye-like screening parameter $\xiscreen$.

Similar to previous works \cite{Boguslavski:2023fdm, Boguslavski:2023alu, Boguslavski:2023jvg}, we introduce time markers that roughly separate the different stages in the bottom-up evolution \cite{Baier:2000sb}. 
The star and circle markers are related to the occupancy $\langle pf \rangle/\langle p\rangle$, with the star placed where it first drops below $1/\lambda$ and the circle placed at its minimum value. Finally, the triangle marker is placed when $P_T/P_L=2$, which signals a system that is close to isotropy.

We will frequently rescale the time in units of the relaxation time,
\begin{subequations}
\begin{align}
    \tau_R=\frac{4\pi\eta/s}{\Teps},\label{eq:relaxation-time}
\end{align}
which depends on the shear viscosity $\eta$ over entropy density $s$, and the effective temperature $\Teps$ defined in Eq.~\eqref{eq:Teps}.
Alternatively, we may use the timescale associated with the bottom-up thermalization process \cite{Baier:2000sb, Boguslavski:2023jvg},
\begin{align}
    \taubmss = \left(\frac{\lambda}{4\pi\NC}\right)^{-13/5}/Q_s\,. \label{eq:taubmss}
\end{align}
\end{subequations}

\subsection{Comparison of Debye-like screening prescriptions}

Let us start by discussing the different Debye-like screening prescriptions
\eqref{eq:different_regularization_matrix_element}.
We find that simulations with these matrix elements lead to quantitatively almost identical results and that most observables differ only at a sub-percent level. In \fig\ref{fig:debye-like-comparison}, we compare observables computed using Debye-like prescription 2 and prescription 1. In particular, we show their ratio for 
several moments of the distribution function and components of the energy-momentum tensor, as a function of time rescaled with the relaxation time $\tau_R$.
We additionally show the Debye mass $m_D$, calculated as in \eq \eqref{eq:debye_mass_general}, and the effective infrared temperature $T_\ast$ given by \eq \eqref{eq:Tstar}.
To estimate the statistical uncertainty coming from the Monte Carlo integration in the evaluation of the collision terms, 
we perform several simulations with identical parameters. 
The error bands in the figure are estimates of the standard deviation of our results.

It is visible in \fig \ref{fig:debye-like-comparison} that the different Debye-like screening prescriptions differ less for small couplings, which is expected due to the leading-order equivalence of the different screening prescriptions, in accordance with our discussion in \se\ref{sec:debye-like-screening}.
However, also for larger values of the coupling (even for $\lambda=20$), the considered observables deviate only at a percent level.
This indicates that small-angle scatterings, in which $|t|\ll s$, strongly dominate the elastic collision term.
This fact is well known and used as the basis for other studies that use the diffusion approximation \cite{Blaizot:2013lga, BarreraCabodevila:2022jhi, Cabodevila:2023htm}. 
Indeed, in \app\ref{app:validity_screening}, we provide numerical evidence that further confirms the dominance of $|t|\ll s$ processes.
The good agreement of these prescriptions highlights the importance of the $t$ channel processes for elastic scatterings.
These phase-space considerations are independent of the coupling and are, therefore, valid also at larger couplings.

Thus, the almost perfect agreement of the observables 
throughout the Debye-like screening prescriptions is a combination of two effects: On the one hand, these screening prescriptions only differ up to $\mathcal O(g)$ for weak couplings, and on the other hand, they are equivalent for small angle scatterings, which are the dominant processes in the relevant kinematic region.

\subsection{Comparison with isoHTL: Pressure ratio}

In contrast to the Debye-like screening prescriptions discussed in the last section, the isoHTL screening leads to significant differences in all considered observables. We will first discuss this at the example of the pressure ratio, and then consider other quantities.%
\footnote{
When numerically evaluating the isoHTL matrix element \eqref{eq:full_isotropic_HTL_matrixelement},
we found that an improved adaptive step size algorithm is required that we describe in detail in \app\ref{app:adative_stepsize}
to obtain a sufficiently smooth time evolution.
}

\begin{figure}
    \centering
    \includegraphics[width=\linewidth]{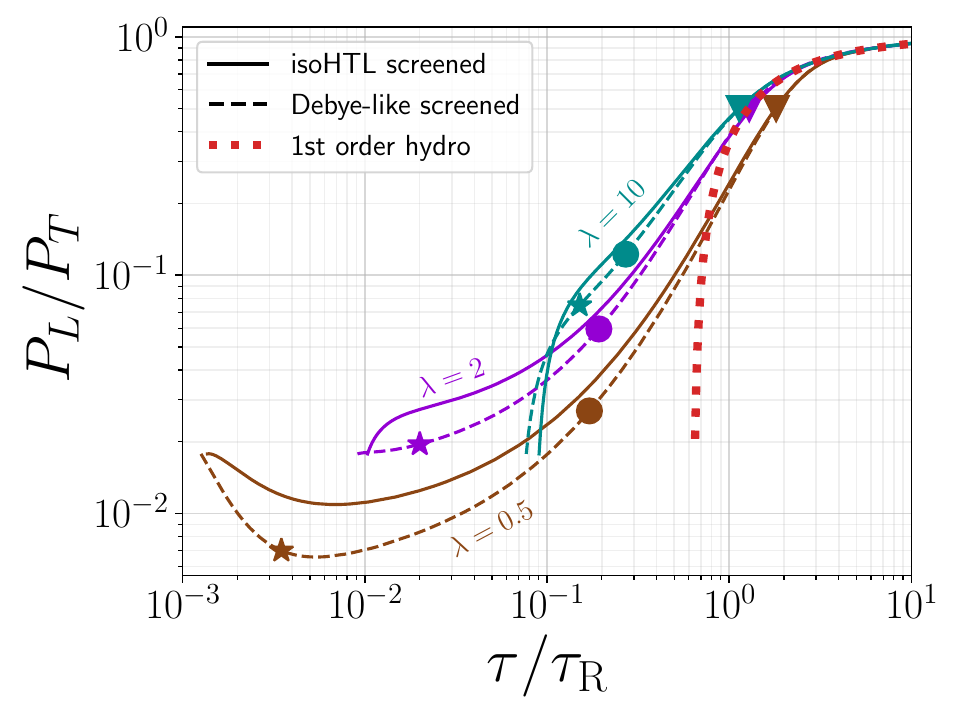}
    \includegraphics[width=\linewidth]{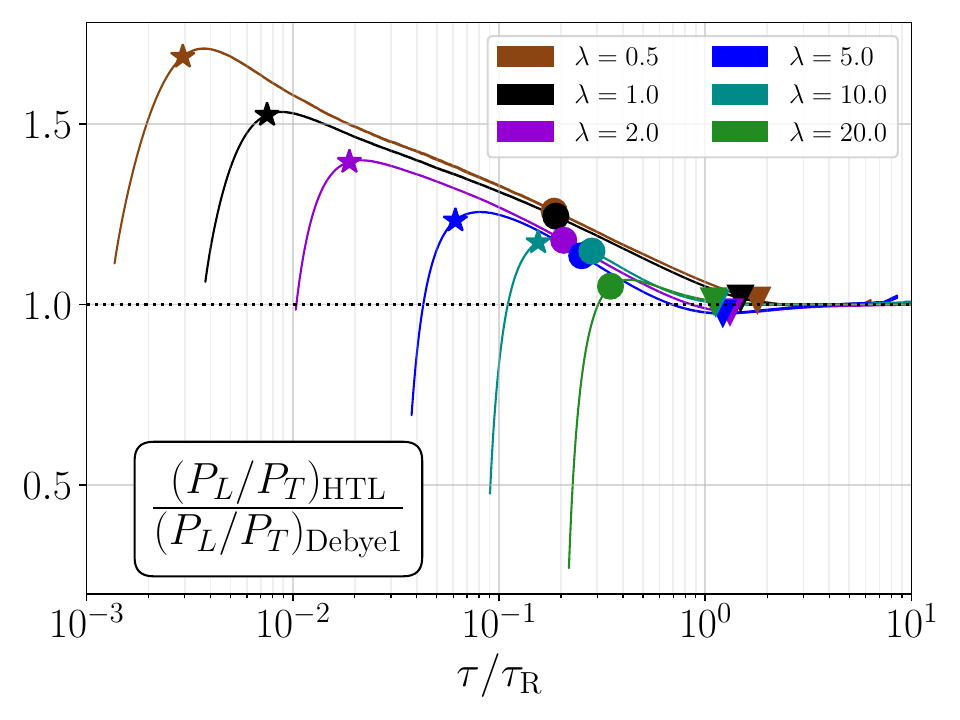}
    \caption{Pressure ratio as a function of time for different couplings (colors). In the top panel, the results using the isoHTL-screened matrix element are shown as solid lines, for the Debye-like screened matrix element as dashed lines. The first-order hydrodynamic estimate is shown as a thick red dotted line. The bottom panel shows the pressure ratio of the isoHTL-screened simulations normalized to the pressure ratio with Debye-like screening. Time is rescaled with the kinetic relaxation time $\tau_R$.
    }
    \label{fig:pressure-ratio}
\end{figure}

The pressure ratio for a fixed value of the coupling $\lambda=0.5$ was already presented in the introduction in \fig\ref{fig:lambda05_runs}. In \fig\ref{fig:pressure-ratio}, we now show the pressure ratio for different couplings (color coded) in the top panel and display the ratio between the isoHTL and Debye-like pressure ratio in the bottom panel. 
We find that the maximum pressure anisotropy is almost halved by employing the isoHTL screening prescription, i.e., the lowest value of $P_L/P_T$ is larger.
The time variable is rescaled with the kinetic relaxation time $\tau_R$ \eqref{eq:relaxation-time}, as is typically done to visualize the universal late-time behavior of the pressure ratio \cite{Romatschke:2017ejr},
\begin{align}
    \frac{P_L}{P_T}=1-\frac{2\tau_R}{\pi \tau}, \label{eq:PL_over_PT_first_order_hydro}
\end{align}
which we also show in the upper panel of \fig\ref{fig:pressure-ratio} as a thick red dashed line.

While both the isoHTL-screened (solid line) and Debye-like screened curves (dashed line, corresponding to \eqref{eq:usual_screened_matrix_element}) start at the same value of the pressure ratio by virtue of the same initial condition, the time evolution with the different screening prescriptions shows clear differences: Simulations with Debye-like screening become more anisotropic as compared to isoHTL screening, which is more pronounced at small values of the coupling $\lambda$.

This can be understood from the observation that while the Debye-like screening prescription \eqref{eq:usual_screened_matrix_element} approximates well the longitudinal momentum transfer, it underestimates transverse momentum broadening as encoded in the jet quenching parameter $\qhat$ \cite{Boguslavski:2023waw}. 
However, transverse momentum broadening
is an essential ingredient in the bottom-up equilibration process \cite{Baier:2000sb}. The Debye-like screening prescription, therefore, leads to less efficient transverse momentum broadening and to a larger anisotropy.

In contrast, the late-time evolution is less sensitive to the screening prescription if the relaxation time in \eq \eqref{eq:relaxation-time} is adjusted accordingly. The required parameter $\eta/s$ 
is obtained such that the curves follow this late-time first-order hydrodynamic estimate \eqref{eq:PL_over_PT_first_order_hydro}. We will study the approach to hydrodynamics and thermalization first and then discuss our extraction procedure of $\eta/s$ together with our numerical results in \se \ref{sec:extract_etas}.

\subsection{Thermalization and hydrodynamization time}
\begin{figure}
    \centering
    \includegraphics[width=\linewidth]{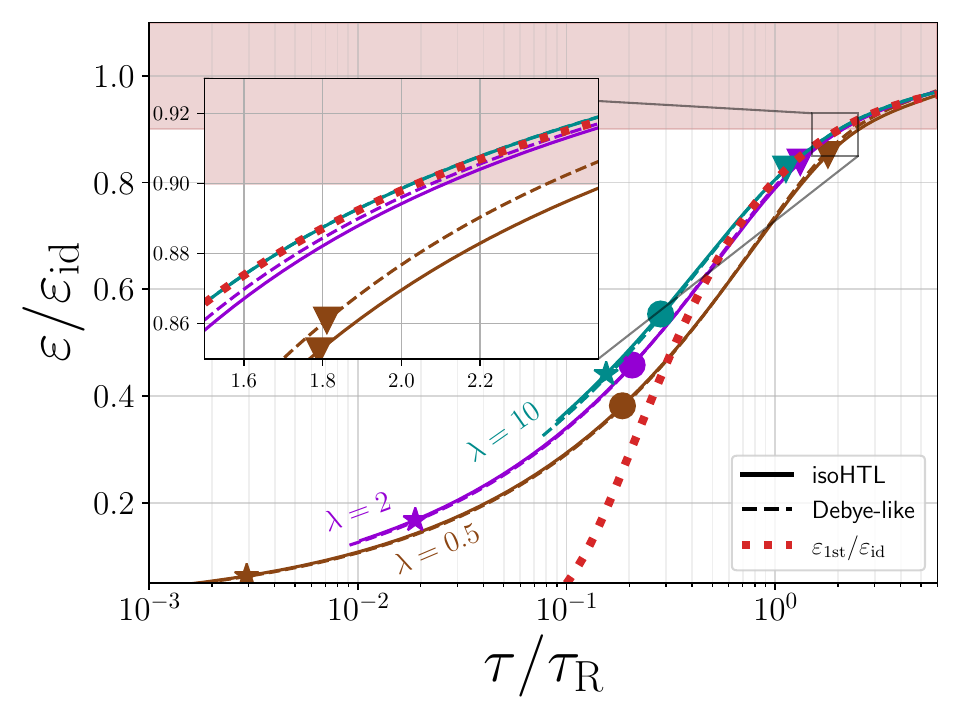}
    \includegraphics[width=\linewidth]{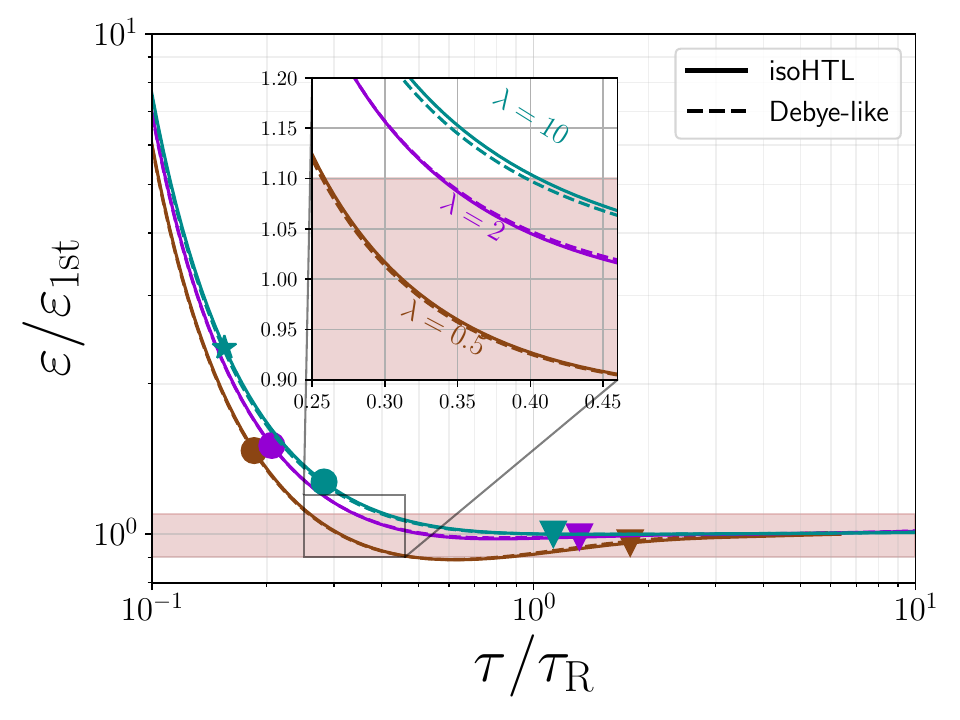}
    \caption{Energy density of the nonequilibrium simulation for different couplings $\lambda$ (color coded) normalized to its ideal (top) and first-order hydrodynamic estimate (bottom). The results using the isoHTL-screened matrix element are shown as solid lines, for the Debye-like screened matrix element as dashed lines. In the top panel, we also depict the first-order hydrodynamic estimate as a thick red dotted line. The shaded region indicates a less than $10$~\% deviation from unity.
    }
    \label{fig:thermalization_and_hydrodynamization}
\end{figure}
Motivated by Ref.~\cite{Kurkela:2018xxd},
we also study the changes in the thermalization and hydrodynamization time scales when using the differently screened matrix elements.
For expanding systems, we define the thermalization time as the time when the energy density of the nonequilibrium kinetic theory simulation gets within $10\%$ of the ideal hydrodynamic estimate for the first time,
\begin{align}
    \left|1-\frac{\varepsilon(\tautherm)}{\varepsilon_{\mathrm{id}}(\tautherm)}\right|
    =0.1\, .
\end{align}
Similarly, the hydrodynamization time is defined as the time when the full kinetic theory result gets within $10\%$ of the first-order hydrodynamic estimate,
\begin{align}
    \left|1-\frac{\varepsilon(\tauhydro)}{\varepsilon_{\mathrm{1st}}(\tauhydro)}\right|=0.1
\end{align}
In ideal and first-order hydrodynamics, the temperature evolution is given by
\begin{align}   
    \Tid(\tau)&=\frac{c_1}{\tau^{1/3}},\\
    \frac{\Tfirst(\tau)}{\Tid(\tau)}&=1-\frac{1}{6\pi}\frac{\tauR}{\tau},
\end{align}
with a constant $c_1$. At the top of \fig \ref{fig:thermalization_and_hydrodynamization}, we show the energy density of our simulation normalized to the energy density of ideal hydrodynamics for different values of the coupling $\lambda$. We observe that both the simple screening results (dashed lines, corresponding to Eq.~\eqref{eq:usual_screened_matrix_element}) and the lines associated with simulation with the isotropic HTL matrix element (solid lines, corresponding to Eq.~\eqref{eq:full_isotropic_HTL_matrixelement}) lie almost on top of each other. This is due to the rescaling of the time using the relaxation time $\tauR$, in which the different values for the shear viscosity $\eta/s$ are already included. 
In particular, we find that thermalization occurs roughly at the same time, at $\tautherm\approx 2\,\tauR$ for $\lambda \gtrsim 2$.

In the bottom panel of \fig\ref{fig:thermalization_and_hydrodynamization} we show the energy density normalized to the first-order hydrodynamic estimate. Similar to the thermalization case, we find that hydrodynamization occurs roughly at the same time when comparing both screening approximations.
The results for different values of the coupling $\lambda$ differ stronger than the different screening prescriptions.
Hydrodynamization occurs roughly at $\tauhydro\approx 0.35\,\tauR$.

These results indicate that the impact of changing the screening prescription on the approach to hydrodynamics and to local thermal equilibrium is mainly captured by the values of the coefficient $\eta/s$.

\subsection{Shear viscosity over entropy density $\eta/s$\label{sec:extract_etas}}

\begin{figure}
    \centering
    \includegraphics[width=\linewidth]{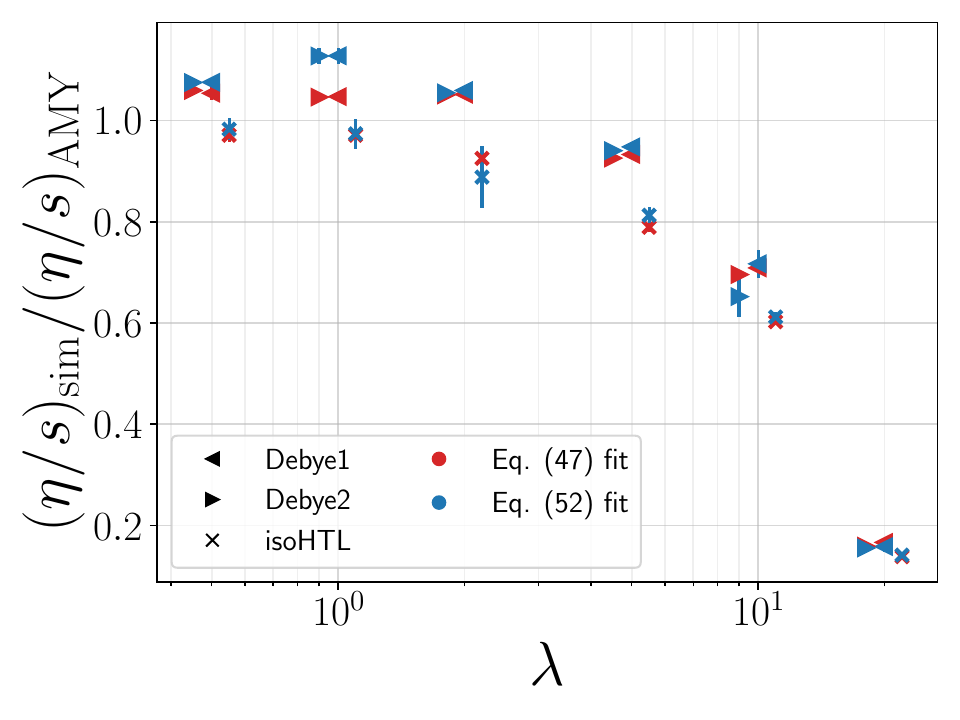}
    \caption{Fitted values for $\eta/s$ from \tab\ref{tab:etas_values}. The simulation and fit results for the different screening methods and fitting formulas are normalized by the NLL pQCD results stated in the table. For visual ease, the markers are slightly shifted around the value of $\lambda$. 
    }
    \label{fig:etas_figure2}
\end{figure}

\subsubsection{Numerical extraction and comparison of $\eta/s$}

As input for the relaxation time $\tauR$ in \eqref{eq:relaxation-time}, we need to obtain the transport parameter $\eta/s$ for every coupling and screening prescription used. This first-order hydrodynamic parameter quantifies the late-time approach to isotropy.
It has been obtained perturbatively in Ref.~\cite{Arnold:2003zc}.

In general, our strategy is to compare the late-time evolution of our numerical simulations with a first-order conformal hydrodynamic system, in which the only medium parameter is the shear viscosity over entropy density $\eta/s$.
There, for a Bjorken expanding system, the pressure ratio is given by Eq.~\eqref{eq:PL_over_PT_first_order_hydro}.
However, this is not the only function that can be used to obtain $\eta/s$. For example, in Ref.~\cite{Kurkela:2018vqr} the matching condition
\begin{align}
    \frac{P_T-P_L}{\varepsilon+P}=2\left(\frac{\eta/s}{\tau \Teps}\right)+4C_2/3\left(\frac{\eta/s}{\tau \Teps}\right)^2,\label{eq:pressure_difference_formula}
\end{align}
is used, which stems from a second-order hydrodynamic formulation. Here, $C_2$ is an additional fitting constant and the pressure $P$ is related to the energy density $\varepsilon$ via $P=\varepsilon/3$ for a conformal system.
We will use both expressions \eqref{eq:PL_over_PT_first_order_hydro} and \eqref{eq:pressure_difference_formula} to obtain the value of $\eta/s$ for a given coupling $\lambda$ and screening prescription. Both equations give similar but not identical results, and their variation provides an estimate of our systematic uncertainty.

In particular, we extract the parameter $\eta/s$ by fitting 
at some late time interval $(t_i,t_f)$ to the pressure anisotropy $P_L/P_T$ of the simulation.
There are two main causes of inaccuracy in this procedure: On the one hand, any fixed-order hydrodynamic formula only describes the system's behavior sufficiently close to thermal equilibrium (in our case at sufficiently late times).
Before that, corrections to it are expected, which will make our numerical fit worse when decreasing $t_i$. On the other hand, our numerical simulations are plagued with discretization artifacts that become worse over time.%
\footnote{For example, due to Bjorken expansion, all characteristic momentum scales decrease, while we use a fixed momentum grid in our simulations.}
Thus, a large final time $t_f$ may worsen the fit.

To obtain the best fit, we, therefore, vary the start ($t_i$) and end times ($t_f$) of our fitting process. We choose as the earliest time for $t_i$ the time when 
the pressure anisotropy is $P_L/P_T=0.5$. 
For each $(t_i, t_f)$ pair, we record the value of $\eta/s$ and its fit error.
Calculating also the gradients $\nabla_{t_i,t_f}\eta/s$ allows us to extract the value where the fit-error times the gradient is the smallest. We then perform this fitting procedure for several simulation runs 
and average the results. By that, we obtain the statistical uncertainty using a simple error of the mean estimate.
We note that the extraction procedure with varying $(t_i, t_f)$ and emerging simulation artifacts may lead to additional systematic errors that we may not sufficiently account for.

The results of this procedure are summarized in \tab\ref{tab:etas_values}.%
\footnote{Note that $\eta/s$ is a medium parameter independent of the specific initial condition. While for $\lambda\geq 1$, we use the same distribution \eqref{eq:initial_cond} as for the expanding simulations, for $\lambda=0.5$ we reduce late-time discretization artifacts by choosing our initial time $Q_s\tau=50$ and initial anisotropy $\xianiso=2$ with the suitable initial amplitude $A(\xianiso{=}2)=0.96789$.}
An immediate observation is that both Debye-like screening prescriptions combined with both fit formulas generally provide similar values for $\eta/s$ for each coupling, and these values are similar to earlier numerical extractions \cite{Keegan:2015avk}. The (systematic) variation from the screening and fitting forms is hereby larger than the uncertainty from the fits. However, the isoHTL values are, in general, about 10\% - 20\% smaller than those from Debye-like screening, and are much closer to the perturbative estimates \cite{Arnold:2003zc}, which are labeled as ``NLL pQCD'' and given by
\begin{align}
    \label{eq:pQCD_NLL}
    \left.\frac{\eta}{s}\right|_{\mathrm{NLL\,pQCD}}=\frac{34.784}{\lambda^2\ln\left[4.789/\sqrt{\lambda}\right]}.
\end{align}
Note that this formula is based on an expansion in inverse logarithms and is thus only valid at small couplings at next-to-leading logarithmic (NLL) accuracy.
The observation that this perturbative estimate is very close to our extracted isoHTL-screened values for small couplings gives us confidence that there are no large systematic biases in our extraction procedure. In general, any such systematic bias would be similar for all screening prescriptions considered in this paper, and we thus expect the decrease of the value of $\eta/s$ for the isoHTL screening prescription to be a robust statement.

\begin{table*}
\begin{ruledtabular}
\begin{tabular}{c c c c c c c c c}
$\lambda$ & Debye1, \eqref{eq:PL_over_PT_first_order_hydro} &  Debye1, \eqref{eq:pressure_difference_formula} & Debye2, \eqref{eq:PL_over_PT_first_order_hydro} &  Debye2, \eqref{eq:pressure_difference_formula} & isoHTL, \eqref{eq:PL_over_PT_first_order_hydro} &  isoHTL, \eqref{eq:pressure_difference_formula} & NLL pQCD\\ [0.5ex]
\hline

$0.5$ & $76.6(9)$ & $78.2(1)$ & $77.0(2)$ & $78.2(1)$ & $70.6(9)$ & $71(2)$ & 72.7\\
$1.0$ & $23.249(6)$ & $25.0(3)$ & $23.228(6)$ & $25.0(3)$ & $21.55(6)$ & $21.6(7)$ & 22.2\\
$2.0$ & $7.50(4)$ & $7.55(4)$ & $7.49(4)$ & $7.517(10)$ & $6.59(9)$ & $6.3(4)$ & 7.13\\
$5.0$ & $1.7037(5)$ & $1.731(5)$ & $1.6909(4)$ & $1.717(1)$ & $1.44(2)$ & $1.48(3)$ & 1.83\\
$10.0$ & $0.5940(6)$ & $0.60(2)$ & $0.5830(7)$ & $0.55(3)$ & $0.505(2)$ & $0.513(8)$ & 0.838\\
$20.0$ & $0.2120(5)$ & $0.20(1)$ & $0.2027(6)$ & $0.20(1)$ & $0.1759(3)$ & $0.180(2)$ & 1.27\\[1ex]
\end{tabular}
\end{ruledtabular}
\caption{Extraction of $\eta/s$ values for various couplings $\lambda$ using the screening prescriptions Debye1 \eqref{eq:usual_screened_matrix_element}, Debye2 \eqref{eq:Debye2_screened_matrix_element}, isoHTL \eqref{eq:full_isotropic_HTL_matrixelement} and the extraction procedures \eqref{eq:PL_over_PT_first_order_hydro} and \eqref{eq:pressure_difference_formula}. The error estimate from our fitting procedure is explained in the text. 
NLL pQCD denotes the weak coupling perturbative expression \eqref{eq:pQCD_NLL} from \cite{Arnold:2003zc}.
}
\label{tab:etas_values}
\end{table*}

In \fig\ref{fig:etas_figure2}, we plot the ratio of our extracted values of $\eta/s$ over the perturbative NLL results from Ref.~\cite{Arnold:2003zc}. We confirm the above-mentioned observations that the $\eta/s$ values for $\lambda \lesssim 1$ for the isoHTL matrix element are very close to the pQCD values, whereas the Debye-like screened matrix elements consistently lead to larger values of $\eta/s$.
At larger couplings, we see increasing discrepancies between the perturbative NLL values and our extracted results where the isoHTL screening continues to yield consistently smaller values than the Debye-like prescriptions.
At such large coupling strengths, we should recall that
the inverse log expansion from Ref.~\cite{Arnold:2003zc} breaks down, and also kinetic theory becomes less accurate.
However, it is interesting and encouraging to note that both screening prescriptions lead to rather similar values even at large couplings.

\begin{table}
\begin{ruledtabular}
\begin{tabular}{c c c}
$\lambda$ & Debye1, \eqref{eq:PL_over_PT_first_order_hydro} &  Debye1, \eqref{eq:pressure_difference_formula} \\ [0.5ex]
\hline
$0.53$ & $70.49(3)$ & $72.4(8)$\\
$2.2$ & $6.39(3)$ & $6.417(4)$\\
$11.2$ & $0.5012(8)$ & $0.507(8)$
\\[1ex]
\end{tabular}
\end{ruledtabular}
\caption{Extraction of $\eta/s$ values for different couplings $\lambda$ and the Debye-like screening \eqref{eq:usual_screened_matrix_element} for the $\eta/s$ extraction procedures \eqref{eq:PL_over_PT_first_order_hydro} and \eqref{eq:pressure_difference_formula}. 
}
\label{tab:etas_values2}
\end{table}

\subsubsection{Simulations with the same $\eta/s$}
\begin{figure}
    \centering\includegraphics[width=\linewidth]{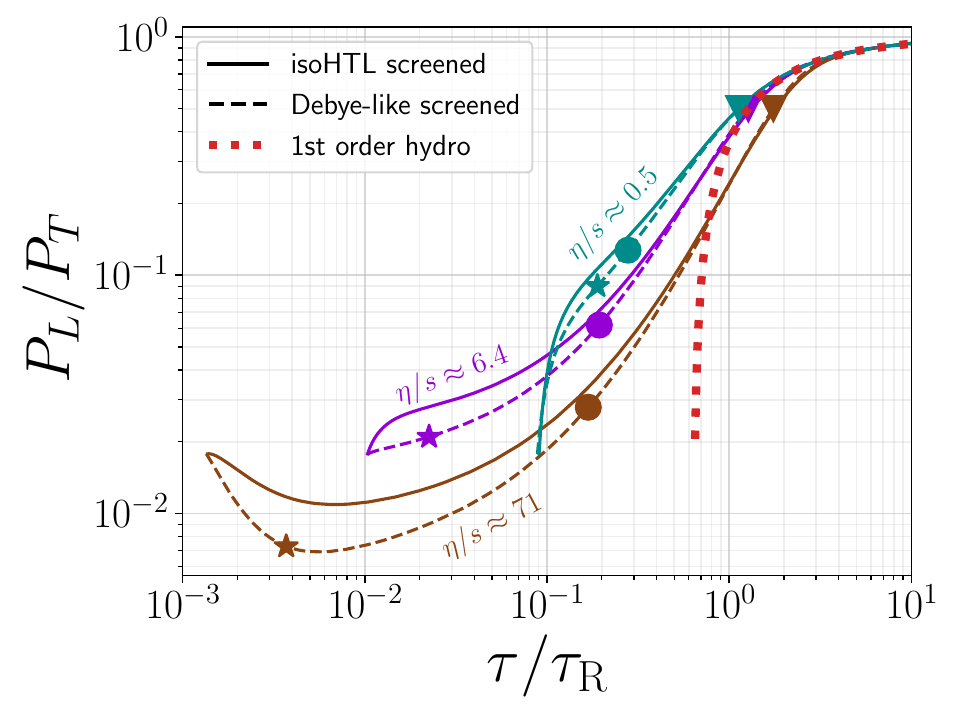}
    \caption{
    Pressure ratio for systems with similar $\eta/s$. The values of $\lambda$ are adjusted between the isoHTL and Debye-like screened simulations to yield a similar value of $\eta/s$.
    }
    \label{fig:pressure_ratio_etas}
\end{figure}
In the previous sections, we have compared the results of Debye-like and isoHTL screening for the same value of the coupling $\lambda$.
We now focus on comparing simulations with the same value of $\eta/s$ to find out whether matching this parameter reduces the deviations between the different screening prescriptions.

In particular, we take the isoHTL screened simulations at couplings $\lambda\in\{0.5,2,10\}$ and compare them to Debye-like screened simulations with couplings $\lambda\in\{0.53,2.2,11.2\}$, which lead to similar values of $\eta/s$ (see \tab\ref{tab:etas_values2}).
This comparison is illustrated in \fig \ref{fig:pressure_ratio_etas} for the example of the pressure ratio $P_L/P_T$ as a function of the rescaled time $\tau/\tauR$. One observes that the late-time behavior of the curves coincides and falls on the universal hydrodynamic curve \eqref{eq:PL_over_PT_first_order_hydro} (red dotted line) as expected from an approach to a hydrodynamical evolution. 

In contrast, the evolution at early times 
differs substantially for the screening prescriptions. 
This observation is very similar to the comparison between screening prescriptions at the same value of the coupling in \fig\ref{fig:pressure-ratio}, and we will study this in more detail in the following. Thus, we conclude that the early-time evolution of the pressure ratio (and similarly of other quantities) is significantly different for the isoHTL screening prescription, even when compared at the same value of $\lambda$ or $\eta/s$.%

\subsection{Impact of screening prescriptions on early-time dynamics}

\begin{figure}
    \centering
    \includegraphics[width=\linewidth]{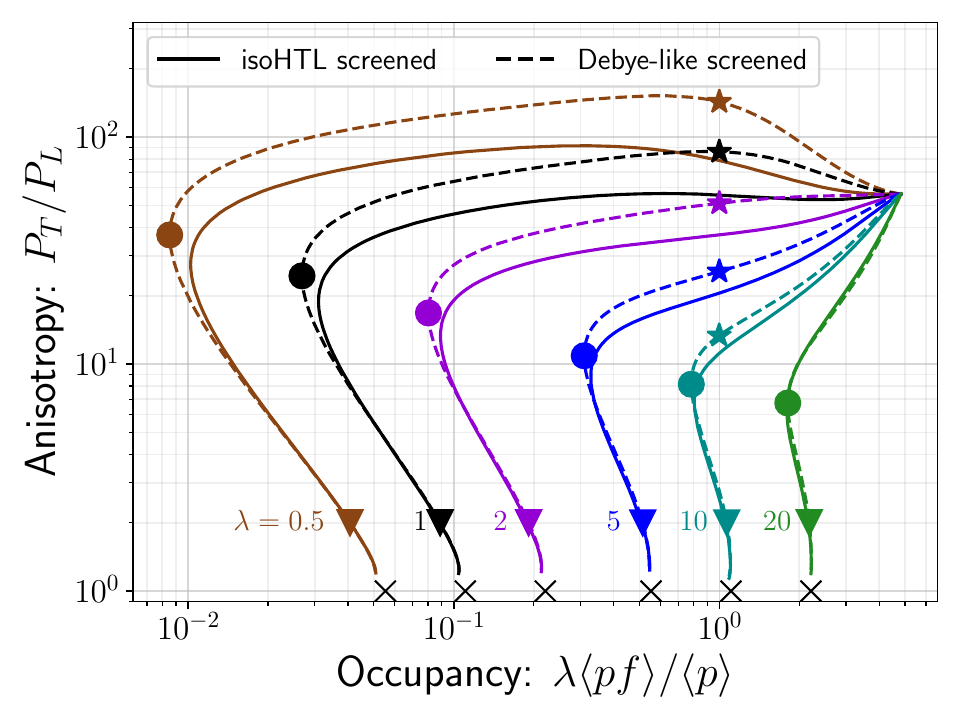}
    \caption{Comparison of simulations with isoHTL (solid and) Debye-like (dashed lines) screening prescriptions in terms of $P_T/P_L$ and occupancy $\lambda \langle pf \rangle/\langle p \rangle$.
    }
    \label{fig:overview-curves}
\end{figure}
\begin{figure*}
    \centerline{% m, lambdaTs, n
        \includegraphics[width=0.33\linewidth]{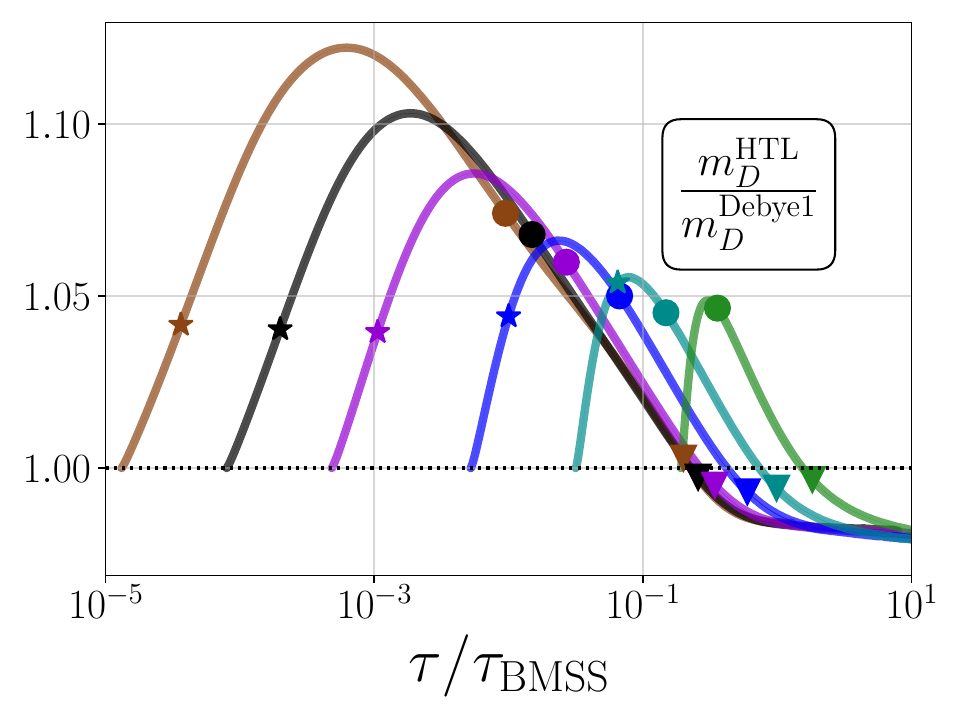}
        \includegraphics[width=0.33\linewidth]{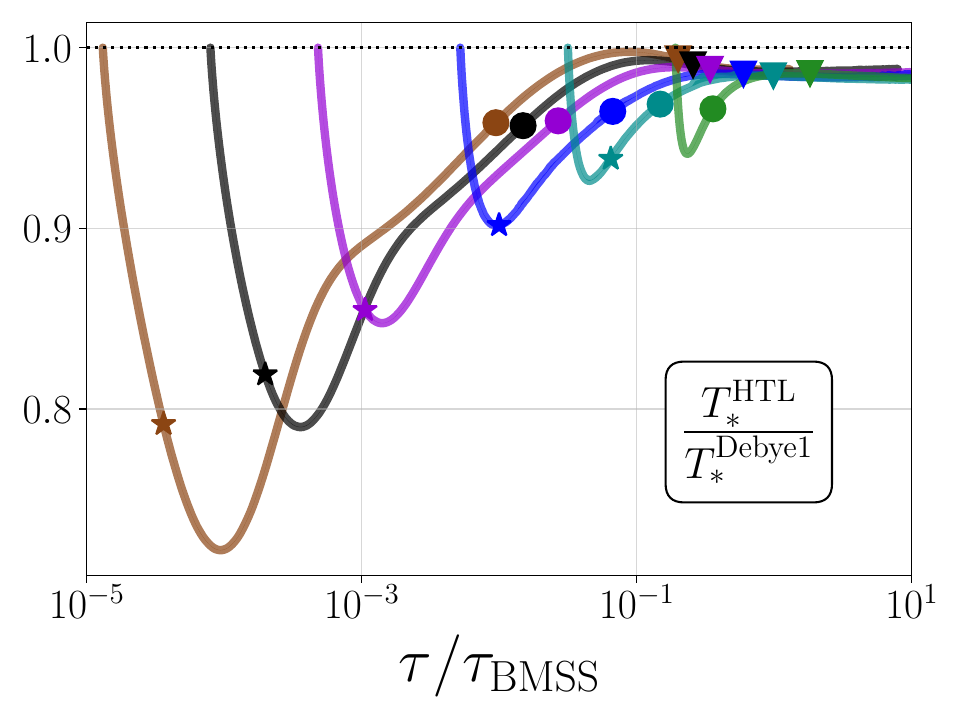}
        \includegraphics[width=0.33\linewidth]{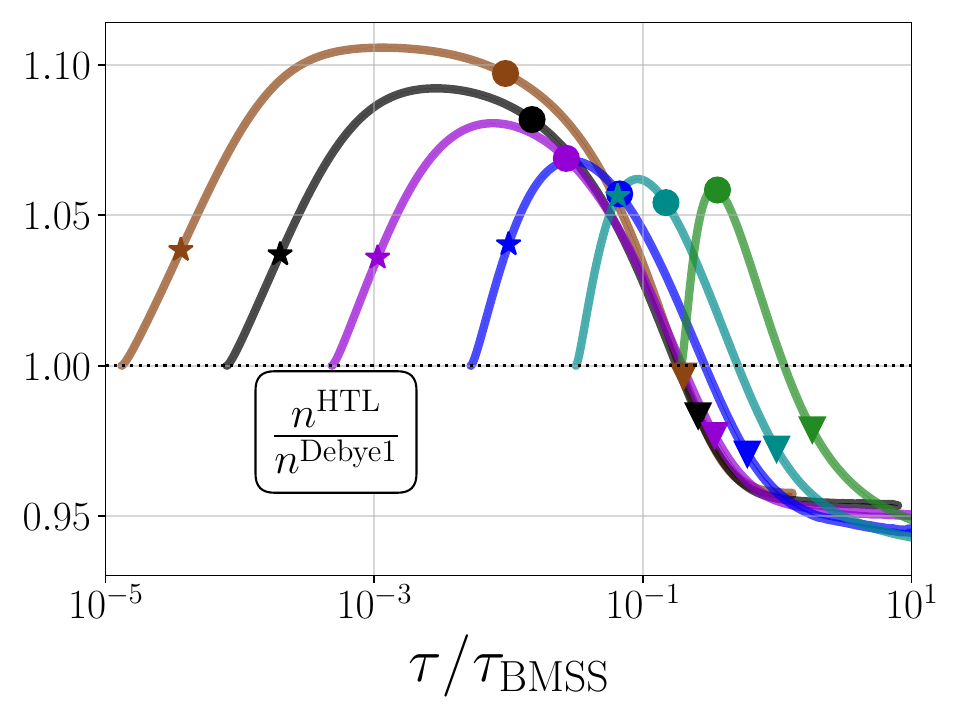}
    }
    \centerline{ %e, PT, PZ
        \includegraphics[width=0.33\linewidth]{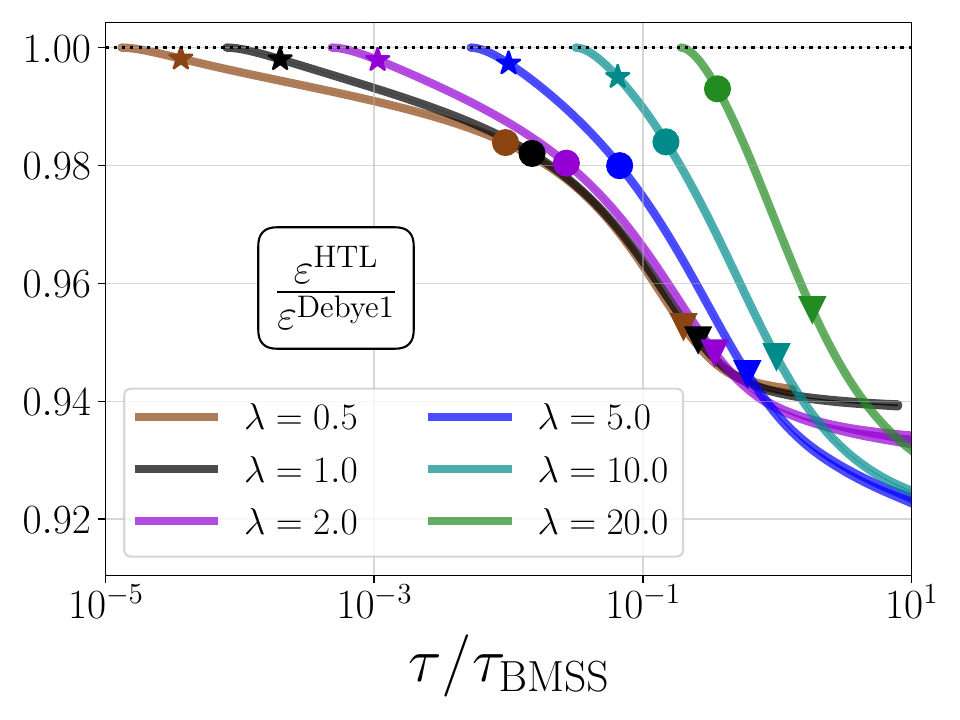}
        \includegraphics[width=0.33\linewidth]{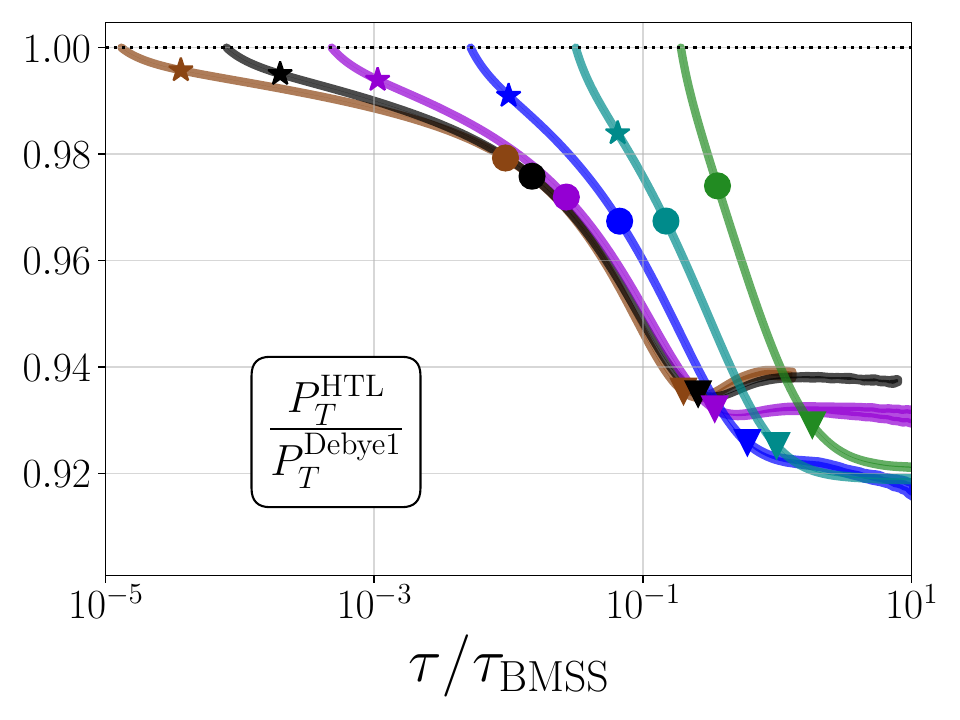}
        \includegraphics[width=0.33\linewidth]{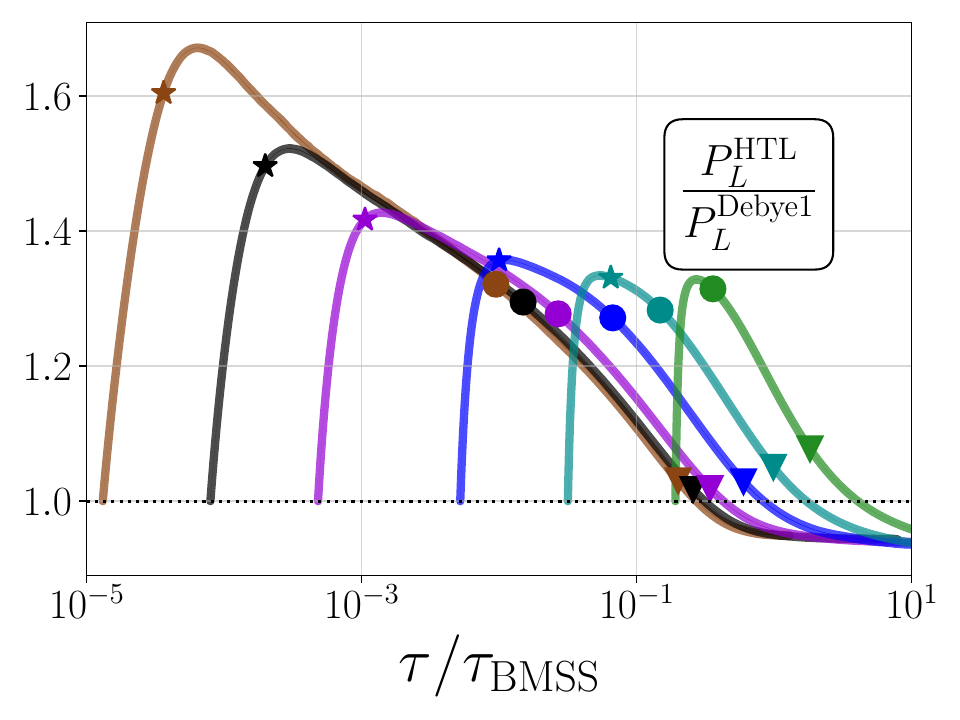}
    }
    \centerline{ % f, pt2, pz2
        \includegraphics[width=0.33\linewidth]{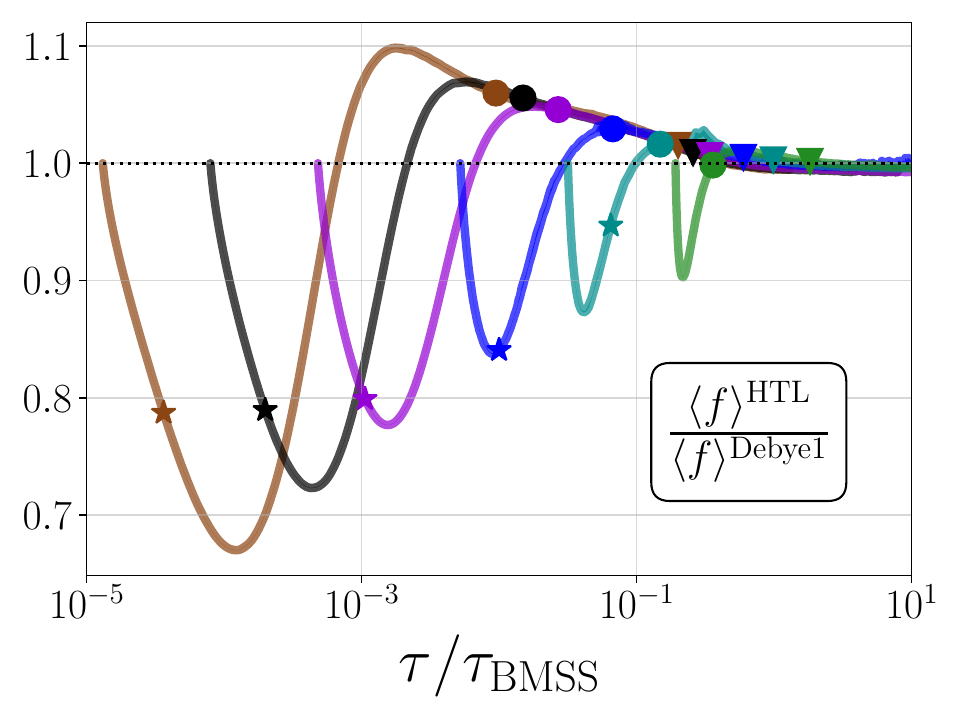}
        \includegraphics[width=0.33\linewidth]{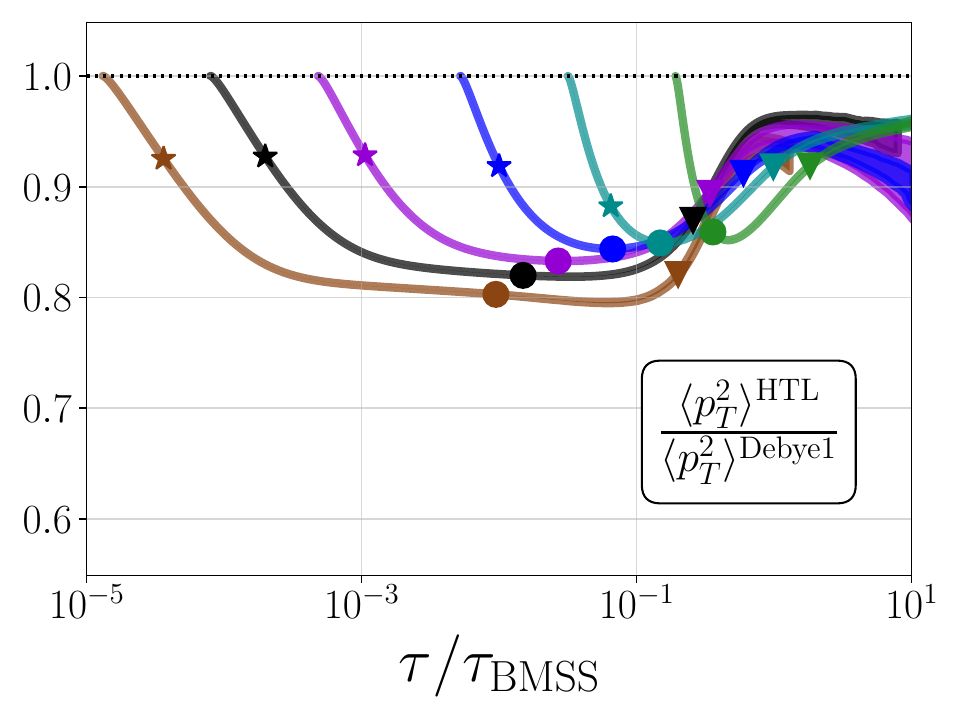}
        \includegraphics[width=0.33\linewidth]{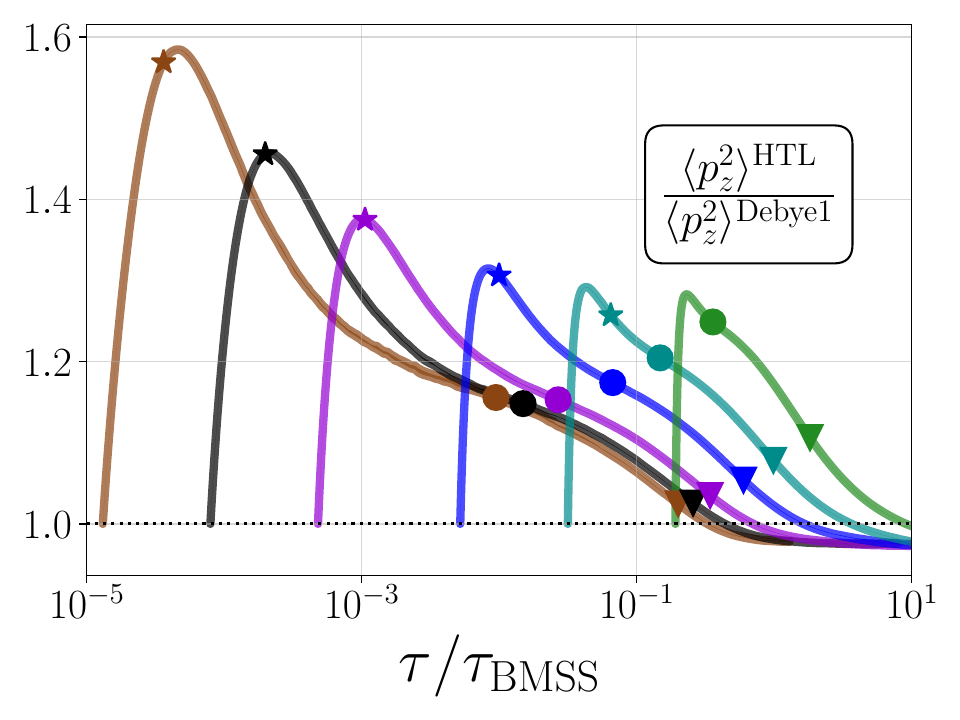}
    }
    \caption{\label{fig:obs_ratios}Ratios of observables obtained from simulations with isoHTL screening divided by those with the Debye-like 1 screening prescription. 
    The time is rescaled with the bottom-up thermalization time \eqref{eq:taubmss} and the error estimates correspond to a statistical average of 5 simulation runs.
    }
\end{figure*}

Visually, the system's evolution to equilibrium is often portrayed in the occupancy-anisotropy plane, as introduced in Ref.~\cite{Kurkela:2015qoa}. 
In \fig\ref{fig:overview-curves}, we reproduce this figure with the Debye-like prescription and additionally include isoHTL-screened simulations. All simulations start with the same initial condition given by \eq \eqref{eq:initial_cond}, shown in the upper-right part of the plot, and then evolve towards thermal equilibrium (black crosses).
We find that the isoHTL matrix element leads to a visibly different evolution, especially during the far-from-equilibrium initial stages before minimal anisotropy is reached (circle marker). These differences are seen throughout a wide range of occupation numbers. One of the effects is that for weak couplings, the system reaches a significantly smaller anisotropy in terms of $P_T/P_L$. Also the minimal occupancy is larger than for Debye-like screening. Nonetheless, we do not find significant qualitative changes to the bottom-up picture of thermalization \cite{Baier:2000sb}.

In \fig \ref{fig:obs_ratios} we study the impact of the isoHTL screening on observables obtained from moments of the distribution function in more detail. 
We plot the ratios of various observables extracted with isoHTL screening over Debye-like screening as functions of rescaled time $\tau/\taubmss$ for different couplings $\lambda$ (color coded), with the thermalization time from Eq.~\eqref{eq:taubmss}.
In particular, we show the Debye mass $m_D$, effective temperature $T_\ast$, particle density $n$, energy density and pressures $\varepsilon$, $P_T$, $P_L$, and moments of the distribution function $\langle f \rangle$, $\langle p_T^2 \rangle$ and $\langle p_z^2 \rangle$. 
We find that most considered quantities show larger deviations between the screening prescriptions for smaller values of the coupling.
This is in line with our expectation that for larger values of the coupling the screening prescriptions differ less, as discussed in \se \ref{sec:ekt_extrapolation_large_couplings}. 
Nonetheless, sizable corrections can be found even at large couplings.

For instance, the Debye mass in the isoHTL simulation is enhanced by almost $15\%$ for $\lambda=0.5$ between the star and circle marker, and by about $5\%$ for $\lambda = 20$. The average occupancy $\langle f \rangle$ 
is reduced 
in the isoHTL simulations by over $30 \%$ at early times for weak couplings, but even for $\lambda=20$ we find a reduction of more than $10\%$. 
In contrast, the particle density is enhanced by $5 - 10 \%$.

However, the largest deviations concern measures of the bulk anisotropy of the plasma.
At late times, the components of the energy-momentum tensor from the isoHTL simulations are about $10\%$ smaller than for the Debye-like screened simulations,
which implies a smaller temperature in the subsequent equilibrium plasma.
In contrast, the longitudinal pressure increases by more than $60\%$ for $\lambda=0.5$, and by more than $30\%$ for $\lambda=10$ and $\lambda=20$ before eventually also decreasing to values similar to $P_T$.
This effect results from the longitudinal pressure being dominated by particles close to the transverse plane that undergo transverse momentum broadening, i.e., in $p_z$ direction. Since isoHTL leads to a more accurate and more efficient transverse momentum broadening than the simple Debye-like screening prescription \cite{Boguslavski:2023waw}, one arrives at the observed higher values of $P_L$. 
Similar effects can be seen in the moments of $f$. Indeed, the anisotropy ratio $\langle p_z^2 \rangle / \langle p_T^2 \rangle$ is also considerably larger for isoHTL simulations than for Debye-like screening and is closer to unity. In particular, $\langle p_z^2 \rangle$ overshoots and $\langle p_T^2 \rangle$ is lower in the HTL cases at early times. This significant reduction of anisotropy at early times is, hence, one of the main qualitative differences between the screening prescriptions. 

\subsection{Evolution of the jet quenching parameter}
\begin{figure}
    \centering
    \includegraphics[width=\linewidth]{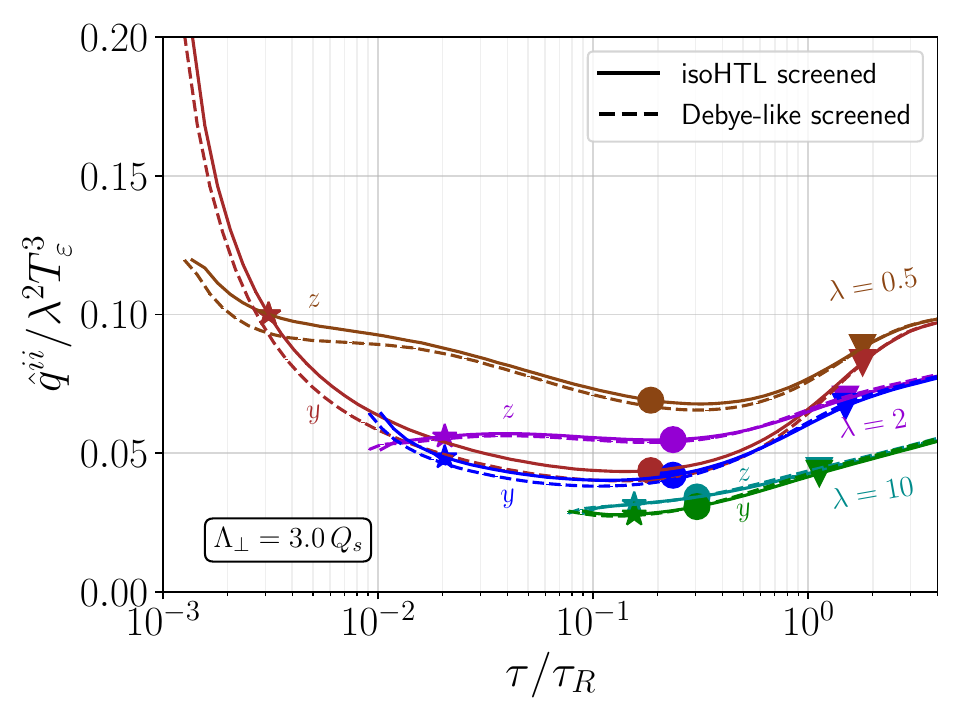}
    \caption{Evolution of $\qhat$ for couplings $\lambda=0.5, 2, 10$. The legend refers to the respective screening prescriptions used in the time evolution.
    }
    \label{fig:qhat}
\end{figure}

Let us finally discuss the change of the screening prescription for the jet quenching parameter $\qhat$, which was studied during the hydrodynamization process recently \cite{Boguslavski:2023alu, Boguslavski:2023waw, Boguslavski:2023jvg}.
It measures the transverse momentum broadening of a highly energetic parton traversing the quark-gluon plasma and can be obtained via \cite{Arnold:2008vd, Boguslavski:2023waw} 
\begin{align}
\hat q^{ij} &= \lim_{p\to\infty}\int_{\substack{\vb k\vb k'\vb p'\\q_\perp < \Lambda_\perp}} q_\perp^i q_\perp^j (2\pi)^4\delta^4(P+K-P'-K') \nonumber \\
&\qquad\times \frac{1}{4 (\NC^2-1)}\,\frac{\left|\mathcal M_{gg}^{gg}\right|^2}{p} f(\vb k)\left(1+ f(\vb k')\right).
\label{eq:qhat_general}
\end{align}
This closely resembles the loss term of the elastic collision kernel \eqref{eq:c22_first}, but still depends on a transverse momentum cutoff $\lperp$. The matrix element appearing in \eqref{eq:qhat_general} is obtained in the limit $p\to\infty$ from Eq.~\eqref{eq:full_isotropic_HTL_matrixelement},
and thus also includes the isoHTL screening prescription. This formulation was also used in our previous work \cite{Boguslavski:2023alu}, but there only Debye-like screening was employed in the Boltzmann equation for the distribution function \eqref{eq:boltzmann_equation}.

We show the time evolution of the momentum broadening coefficient along the beam axis $\qhat^{zz}$ and in the transverse direction $\qhat^{yy}$ in \fig\ref{fig:qhat} for different couplings and screening prescriptions of the plasma.
Our results are quantitatively close to those reported in Ref.~\cite{Boguslavski:2023alu}; the different screening prescriptions have, thus, only a small influence on the time evolution and value of the jet quenching parameter $\qhat$.

\section{Conclusions\label{sec:conclusions}}

In this paper, we studied different screening prescriptions in QCD kinetic theory simulations of the thermalization and hydrodynamization process in heavy-ion collisions for a gluon plasma. 

In particular, we have investigated different ways to include the typically used Debye-like isotropic screening approximation in the elastic collision kernel \eqref{eq:different_regularization_matrix_element}, and the effect of using an isotropic HTL screening (isoHTL), given by \eq \eqref{eq:full_isotropic_HTL_matrixelement}. We find that all simple Debye-like screening prescriptions lead to a very similar time evolution of the plasma, measured in various moments of the distribution function. In contrast, the more accurate isoHTL screening prescription leads to significant changes in these quantities, in particular at small couplings or early times.
For example, at the same value of the coupling, the maximum pressure anisotropy $P_L/P_T$ is reduced by up to $70\%$ for $\lambda=0.5$ and $30\%$ for $\lambda=20$ 
for the isoHTL screening prescription. We arrived at a similar conclusion for typical longitudinal and transverse momenta. This is seen when comparing simulations with the same coupling $\lambda$ as well as with the same shear viscosity over entropy density ratio $\eta/s$.
This indicates that the treatment of the screening prescription may lead to a significantly different evolution at early times, with a substantial decrease in the maximum plasma anisotropy.

We have also investigated the effect of these screening prescriptions on the late-time evolution,
which describes the approach to hydrodynamics.
For this, the transport parameter $\eta/s$ is essential. We found that the isoHTL screening prescription consistently leads to smaller values of $\eta/s$ by about $10 \% $ - $20 \% $ for the same values of the coupling, and reproduces results from perturbative QCD at NLL accuracy \cite{Arnold:2003zc} for sufficiently small $\lambda$.

We also studied the influence 
of screening prescriptions in the dynamics of the plasma on the jet quenching parameter $\qhat$ and observed only a small impact on its value and anisotropy.
Similarly, we did not find any substantial differences in the thermalization process of isotropic systems. Therefore, the Debye-like screening prescription for the evolution of the plasma provides a good approximation to the more general isoHTL screening in these cases.

So far, we still rely on isotropic screening approximations in the Bjorken expanding anisotropic case due to the lack of proper treatment of plasma instabilities that appear in anisotropic systems. 
Generalizing our approach to an anisotropic screening prescription is left as a future task.
It would also be interesting to study the implications of such a prescription on the preequilibrium evolution, particularly far from equilibrium, where we found the quantitatively largest differences.

Additionally, QCD kinetic theory has been used to study the impact of the anisotropies in the initial nonequilibrium system on hard probes moving through the preequilibrium plasma \cite{Boguslavski:2023alu, Boguslavski:2023fdm, Du:2023izb, Zhou:2024ysb} 
and photon or dilepton production \cite{Coquet:2021lca, Hauksson:2023dwh, Garcia-Montero:2023lrd, Giacalone:2019ldn, Garcia-Montero:2024lbl}. Since all previous results of QCD kinetic theory simulations were obtained using the simple Debye-like isotropic screening, it would be interesting to study how these results are modified when using a more general screening prescription such as isoHTL.
%%%%%%%%%%%%%%%%%%%%%%%%%%%%%%%%%%%%%%%%%%%%%%%

\begin{acknowledgments}
We would like to thank Aleksi Kurkela for his assistance and invaluable insights in using the EKT code, and Tuomas Lappi, Aleksas Mazeliauskas and Jarkko Peuron for valuable discussions and collaboration on related projects.
This work is supported by the Austrian Science Fund (FWF) under Grant DOI 10.55776/P34455. FL is additionally supported by the Doctoral Program W1252-N27 Particles and Interactions [Grant DOI: 10.55776/W1252]. The results in this paper have been achieved in part using the Vienna Scientific Cluster (VSC), project 71444.
\end{acknowledgments}

\appendix

\section{Discretization\label{app:discretization}}

In kinetic theory, all medium information is encoded in the distribution function $f(\vb p)$. In our case, it depends only on two parameters, the magnitude of the momentum $p$ and its polar angle $\theta$.
For our numerical implementation, instead of the distribution function $f(\vb p)$, we follow \cite{AbraaoYork:2014hbk} and store the quantity 
\begin{align}
    n_{ij}=\lambda\int\frac{\dd[3]{\vb p}}{(2\pi)^3}f(\vb p) w_i(p)  \tilde w_j(\cos\theta),\label{eq:def_nij}
\end{align}
with the piecewise linear wedge functions $w$, defined for the momentum case as
\begin{align}
    w_i(p)=\begin{cases}
        \frac{p-p_{i-1}}{p_i-p_{i-1}}, & p_{i-1} < p < p_i\\
        \frac{p_{i+1}-p}{p_{i+1}-p_{i}}, & p_{i} < p < p_{i+1}\\
        0& \text{else,}
    \end{cases}
\end{align}
and for the polar angle similarly with the grid points substituted by $p_i \to \cos\theta_i$. The distribution function $f(\vb p)$ can be recovered from $n_{ij}$ by linear interpolation between the grid points, where it can be obtained by the simple relation
\begin{align}
		f(p_i,\cos\theta_j)=\frac{(2\pi)^3n_{ij}}{\lambda p_i^2 4\pi \Delta V^p_i\Delta V^\theta_j},
	\end{align}
 with the volume factors defined via
 \begin{align}
     \Delta V_i^p=\int_{-\infty}^\infty\dd{x}w_i(x),
 \end{align}
 and similarly for the polar angle.
 We have checked that our results do not depend on our discretization parameters, which are the number of momentum and angular bins, as well as the minimum and maximum grid momentum. 

For the overoccupied isotropic systems in \se \ref{sec:thermalization_isotropic}, we used 300 momentum grid points and grid boundaries $(\pmin,\pmax)=(0.01,10)$ for $\lambda=10$, $(0.01,12)$ for $\lambda=2$, and $(0.01,25)$ for $\lambda=0.5$. For the underoccupied isotropic system, we used 500 grid points and $(0.03,100)$ as grid boundaries. For the expanding systems in \se \ref{sec:results}, we used a $(p, \cos\theta)$ grid with $(200, 180)$ points and $(\pmin,\pmax)=(0.03,8)$.

\section{Kinematic considerations\label{app:kinematic-considerations}}

In this Appendix, we list several kinematic considerations that are useful for our analysis.

\subsection{Why $s$ is always the largest Mandelstam variable}
The Mandelstam variable $s$ corresponds to the square of the center-of-mass energy. Since all Mandelstam variables are Lorentz invariant, we may compute them in a boosted frame as well.

For simplicity, we choose the center-of-mass frame, such that
\begin{subequations}
\begin{align}
    P&=(p,p,0,0),& K&=(p,-p,0,0)\\
    P'&=(p, p\,\vb e_x + \vb q), & K'&=(p,-p\,\vb e_x-\vb q).
\end{align}
\end{subequations}
Note that the energy of each particle separately is conserved in an elastic scattering process in this frame.

From $P'^2=0$ we can compute $2pq_x+q^2=0$, or
\begin{subequations}
\begin{align}
    q_x^2+2pq_x+q_y^2&=0,\\
    q_x&=-p\pm \sqrt{p^2-q_y^2}.
\end{align}
\end{subequations}
Thus, $|q_y| \leq p$, and $|q_x| \leq 2p$. Additionally,
\begin{align}
    q_x=-\frac{q^2}{2p}.
\end{align}
The Mandelstam variable $t$ is given by $t=-q^2$, which is thus bounded by $|t| \leq 4p^2=s$.
Therefore, $s$ is always the largest Mandelstam variable, and no $s$-channel process will need to be screened, as explained in the main text and in Ref.~\cite{Arnold:2002zm}.

\subsection{Kinematic considerations: From $Q^2$ small follows $q,\omega$ small}

In this subsection, we show that the requirement $0<-t\ll s$, or $Q^2\ll -(P+K)^2$ implies that all components of $Q$, i.e., $|\omega|$ and $q$, are small.
For that, we consider a plasma with typical excitations at momentum $T$,
\begin{subequations}
\begin{align}
    P&=T(1,1,0,0), & K&=T(1, \vec n), \\
    P'&=(T+\omega,T+q_x,q_y,q_z), & K'&=(T-\omega,T\vec n - \vec q),
\end{align}
\end{subequations}
with $\vb n$ being a unit vector in the $x-y$ plane, i.e., $n_x^2+n_y^2=1$ and $n_z=0$.
The condition $|t|=|Q^2|\ll s$ is equivalent to
\begin{align}
    q^2-\omega^2\ll 2T^2(1-n_x),\label{eq:mandelstam_requirement}
\end{align}
where we used that $q^2 > \omega^2$.

We show in the following that Eq.~\eqref{eq:mandelstam_requirement} implies $|\vb q| \gtrsim |\omega| \ll T$.
The case $|\vb q| \approx T$ would require $|\omega| \approx T$ for Eq.~\eqref{eq:mandelstam_requirement} to be fulfilled. We will show in the following that this is kinematically forbidden, and thus $|\vb q|\ll T$.

We thus assume $|\omega|\approx T$ and introduce a parameter $\alpha \ll 1-n_x$ to rewrite Eq.~\eqref{eq:mandelstam_requirement} to $q^2-\omega^2=\mathcal O(\alpha^2T^2)$.
From $P'^2=0$ we can 
obtain an expression for $q_x = \omega + \mathcal O(\alpha^2 T)$.
Similarly, from $K'^2=0$, we obtain $q_y=\frac{1-n_x}{n_y}\omega + \mathcal O(\alpha^2T)$. Inserting back into Eq.~\eqref{eq:mandelstam_requirement} yields an equation for $q_z$,
\begin{align}
    q^2-\omega^2 = \frac{1-n_x}{1+n_x}\omega^2+q_z^2+\mathcal O(\alpha^2 T^2)=0,
\end{align}
which has no solution for $q_z^2 > 0$. Therefore, the assumption $q^2\approx\omega^2\approx T^2$ leads to a contradiction, and we have shown that $|\vb q|, |\vb \omega| \ll T$.

\subsection{Coordinate system and Mandelstam variables}
The elastic collision term, after symmetrization (see, e.g., \cite{AbraaoYork:2014hbk}) features an integration over $\vb p$, $\vb k$, and $\vb q$. Similarly to Ref.~\cite{Arnold:2003zc}, we perform the integrals over $\vb p$ and $\vb k$ in a frame, in which $\vb q$ defines the $z$-direction with the original $z$-axis lying in the $xz$ plane. Contrary to Ref.~\cite{Arnold:2003zc}, we do not require $\vb p$ to be in the $xz$ plane, since we also integrate over $\vb p$. This is also in contrast to earlier work for the evaluation of $\qhat$ in Ref.~\cite{Boguslavski:2023waw}. 
We choose as integration variables the exchange momentum $\vb q$, parametrized by its magnitude $q$, polar and azimuthal angle $\cos\theta_q$, and $\phi_q$,
the exchange energy $\omega$, the magnitude of the vectors $p$ and $k$, and the azimuthal angles of $\vb p$ and $\vb k$ in a frame, in which $\vb q$ points in the $z$ direction and the original $z$ axis lies in the $xz$ plane. This fixes the kinematics completely. We denote these azimuthal angles by $\theta_{qp}$ and $\theta_{qk}$. Energy conservation results in the kinematic conditions
\begin{align}
    |\omega| < q, && p > \frac{q-\omega}{2}, && k > \frac{q+\omega}{2}.
\end{align}

In terms of these integration variables, we parameterize the relevant vectors using these variables as
\begin{subequations}\label{eq:external_momenta}
\begin{align}
    Q&=P'-P=\begin{pmatrix}
        \omega\\0\\0\\q
    \end{pmatrix}, \\
    P&=p\begin{pmatrix}
       1\\\sin\thetaqp\cos\phiqp\\ \sin\thetaqp\sin\phiqp \\ \cos\thetaqp  
    \end{pmatrix}, \\
    K&=k\begin{pmatrix}
        1\\ \sin\thetaqk\cos\phiqk \\ \sin\thetaqk \sin\phiqk \\ \cos\thetaqk
    \end{pmatrix}.
\end{align}
\end{subequations}
with the angles $\thetaqk$ and $\thetaqp$ given by
\cite{Arnold:2003zc}
\begin{subequations}
\begin{align}
    \cos\thetaqk&=\frac{\omega}{q}-\frac{t}{2kq},\\
    \cos\thetaqp&=\frac{\omega}{q}+\frac{t}{2pq}.
\end{align}
\end{subequations}
In terms of the integration variables, the Mandelstam variables needed for the matrix elements are given by
\begin{align}
    t &=\omega^2-q^2, \label{eq:mandelstam_t_explicit}\\
\begin{split}
s &= -\frac{t}{2q^2}\Big((p+p')(k+k')+q^2\\
	&\qquad\qquad-\sqrt{(4pp'+t)(4k'k+t)}\cos(\phiqk-\phiqp)\Big),
\end{split} \label{eq:mandelstam_s_explicit}\\
\begin{split}
u &= \frac{t}{2q^2}\Big((p+p')(k+k')-q^2\\
	&\qquad\qquad-\sqrt{(4pp'+t)(4k'k+t)}\cos(\phiqk-\phiqp)\Big),
\end{split}\label{eq:mandelstam_u_explicit}
\end{align}

\section{Scattering with soft-gluon exchange\label{app:scattering_soft_gluon_exchange}}
\begin{figure}
    \centering
    \includegraphics[width=\linewidth]{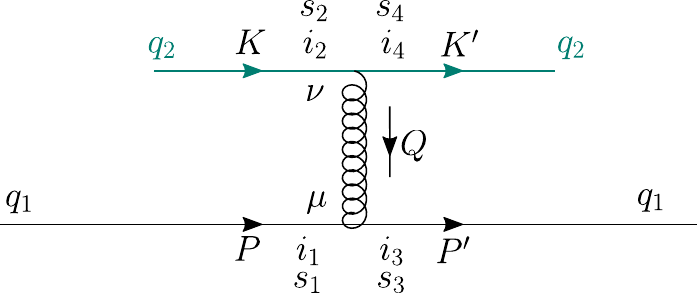}
    \caption{Feynman diagram of quark scattering with two quark flavors with momenta $q_1$ and $q_2$. For illustration, the color, spin and Lorentz indices are shown explicitly.}
    \label{fig:quarkscattering}
\end{figure}
In this Appendix, we show explicitly that quark or gluon scatterings with soft-gluon exchange are equivalent to scalar scatterings as in \eq \eqref{eq:scalar_quark_result} up to $\mathcal O(Q/P)$. Or, put differently, that for soft-gluon exchange the matrix elements are at leading order independent of the spin of the hard particles.

\subsection{Quark scattering with soft momentum transfer}
First, we calculate the square of the matrix element for quark scattering with a general internal gluon propagator.
The amplitude for quark scattering of two different flavors with incoming momenta $P$ and $K$ (spin states $s_1$, $s_2$ and colors $i_1$, $i_2$) to the outgoing momenta $P'$ and $K'$ (spin states $s_3$ and $s_4$, colors $i_3$ and $i_4$), as depicted in \fig\ref{fig:quarkscattering}, is given by%
\footnote{The Feynman rules can be found in any standard QFT textbook, e.g. see \cite{Srednicki:2007qs} for the mostly plus metric convention.}
\begin{align}
	i\mathcal M^{s_1i_1,s_2i_2}_{s_3i_3,s_4i_4} &=
	-g^2\left(\bar u^{s_3}(P')\gamma^\mu u^{s_1}(P)\right)\left(\bar u^{s_4}(K')\gamma^\nu u^{s_2}(K)\right) \nonumber\\
    &\qquad\times \left(-iG_{\mu\nu}^{ab}(Q)\right)t^a_{i_3i_1}t^b_{i_4i_2}\,.
\end{align}
To calculate the square $\left|\mathcal M\right|^2 = \sum_{s_j, i_j} \left(\mathcal M^{s_1i_1,s_2i_2}_{s_3i_3,s_4i_4}\right)^\ast \mathcal M^{s_1i_1,s_2i_2}_{s_3i_3,s_4i_4}$, we need to sum over the color indices and use 
\begin{align} \left(t^a_{i_3i_1}t^b_{i_4i_2}\right)^\ast t^c_{i_3i_1}t^d_{i_4i_2}=\Tr(t^at^c)\Tr(t^bt^d)=n_F^2\delta^{ac}\delta^{bd},
\end{align}
where $n_F=1/2=C_FN_C/d_A$ is the index of the fundamental representation of $SU(\NC)$. Summing over all spins, we obtain
\begin{align}
\begin{split}
	\left|\mathcal M\right|^2 &=\sum_{s_1s_2s_3s_4} n_F^2\delta^{ac}\delta^{bd}g^4G_{\mu\nu}^{ab}(Q)\left(G_{\rho\sigma}^{cd}(Q)\right)^\ast \\
	&\times \left(\bar u^{s_3}(P')\gamma^\mu u^{s_1}(P)\right)\left(\bar u^{s_1}(P)\gamma^\rho u^{s_3}(P')\right)\\
	&\times \left(\bar u^{s_4}(K')\gamma^\nu u^{s_2}(K)\right)\left(\bar u^{s_2}(K)\gamma^\sigma u^{s_4}(K')\right)
 \end{split}
\end{align}
With the identity $\sum_{s_1}u_i^{s_1}(P)\bar u^{s_1}_j(P)=-P_\mu(\gamma^\mu)_{ij}=-\slashed{P}_{da}$, we obtain
\begin{align}
	\left|\mathcal M\right|^2
    &=n_F^2g^4G^{ab}_{\mu\nu}(Q)\left(G^{ab}_{\rho\sigma}(Q)\right)^\ast\\
    &\quad\times\Tr(\gamma^\mu\slashed{P}\gamma^\rho\slashed{P'})\Tr(\gamma^\nu\slashed{K}\gamma^\sigma\slashed{K'})\nonumber.
\end{align}
Using
\begin{align}
	\Tr(\slashed{P}\gamma^\mu\slashed{P'}\gamma^\nu)=4(P^\mu P'{}^\nu + P^\nu P'{}^\mu - g^{\mu\nu}P\cdot P'),
\end{align}
we finally arrive at
\begin{align}
	\begin{split}
	\left|\mathcal M\right|^2&=16 n_F^2g^4 G^{ab}_{\mu\nu}(Q)\left(G^{ab}_{\rho\sigma}\right)^\ast(Q)\\
    &\qquad\times\left[P^\mu P'{}^\rho + P^\rho P'{}^\mu - g^{\mu\rho}P\cdot P'\right]\\
	    &\qquad\times\left[K^\nu K'{}^\sigma + K^\sigma K'{}^\nu - g^{\nu\sigma}K\cdot K'\right].
	\label{eq:quark_scattering_general_matrix_element}
	\end{split}
\end{align}
This expression is independent of the precise form of the gluon propagator $G$. 
Note that all terms in the propagator $G_{\mu\nu}$ proportional to $Q_\mu$ or $Q_\nu$ (which we argue in \se\ref{sec:isoHTLgeneral} are the terms that depend on the specific gauge choice) do not contribute, i.e.,
\begin{align}
    &Q_\mu\left[P^\mu(P+Q)^\nu+ P^\nu(P+Q)^\mu-g^{\mu\nu} P\cdot Q\right]\\
    &=P^\nu \left(2(Q\cdot P)+Q^2\right)=P^\nu(P+Q)^2=P^\nu P'^2= 0.\nonumber
\end{align}

To expand for small $Q$, we use $P'=P+Q$, $K'=K-Q$. 
To lowest order,
the only $Q$-dependence remains in the propagator and we obtain
\begin{align}
	|\mathcal M|^2&=16 n_F^2g^4G_{\mu\nu}G^\ast_{\rho\sigma}\, 4P^\mu P^\rho K^\nu K^\sigma\left(1+\mathcal O\left(\frac{Q}{P}\right)\right).
\end{align}
To the lowest order in $Q$, this is equivalent to 
\begin{align}
	|\mathcal M|^2&=4 n_F^2g^4G_{\mu\nu}G^\ast_{\rho\sigma}\\
    &\qquad\times(P+P')^\mu (P+P')^\rho (K+K')^\nu (K+K') ^\sigma \nonumber\\
    &\qquad\times\left(1 + \mathcal O(Q/P)\right)\nonumber,
\end{align}
which at the same order
is the same as the scalar quark result \eqref{eq:scalar_quark_result}.

\subsection{Gluon scattering with soft momentum transfer}
After having considered the case of quark scattering, we turn to gluon scattering. In this subsection, we will show that in the soft limit, medium effects in gluon scattering enter in the same way.

\begin{figure}
    \centering
    \includegraphics[width=0.6\linewidth]{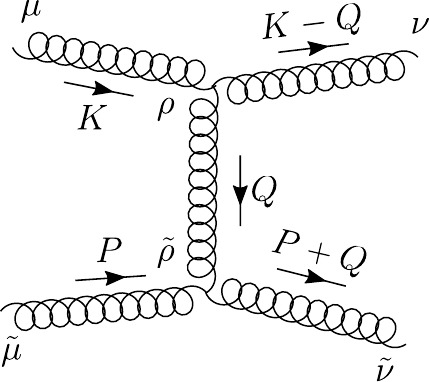}
    \caption{t-channel diagram for gluon scattering.
    }
    \label{fig:gluonscattering}
\end{figure}
Gluon-gluon scattering is depicted pictorially in \fig\ref{fig:gluonscattering}. The dominant process for screening effects is the $t$-channel, where $|t|\ll s$, or---equivalently---the $u$-channel, which can be readily obtained by exchanging $u\leftrightarrow t$ in the collision kernel.

Using the Feynman rules, we can write (suppressing color factors)
\begin{align}
    \mathcal M^{\mu\tilde\mu\nu\tilde\nu}&\propto \left[g^{\mu\rho}(K+Q)^\nu+g^{\rho\nu}(K-2Q)^\mu+g^{\nu\mu}(Q-2K)^\rho\right]\nonumber\\
    &\times \left[g^{\tilde\mu\tilde\nu}(2P+Q)^{\tilde\rho}+g^{\tilde\nu\tilde\rho}(-P-2Q)^{\tilde\mu}+g^{\tilde\rho\tilde\mu}(Q-P)^{\tilde\nu}\right]\nonumber\\
    &\qquad\times G_{\rho\tilde\rho}(Q).\label{eq:gluonscattering-general}
\end{align}
In the soft limit, for small $Q$, we can, to a first approximation, neglect all $Q$ dependence except for the propagator. Additionally, when summing over polarizations, we need to contract the ``same'' external lines $\left(\mathcal M^{\mu\tilde\mu\nu\tilde\nu}\right)^\ast \mathcal M_{\mu\tilde\mu\nu\tilde\nu}$, and then the first line simplifies to
\begin{align}
\begin{split}
    \left[g^{\mu\rho}K^\nu+g^{\rho\nu}K^\mu-2g^{\nu\mu}K^\rho\right]\\
    \times\left[g_{\mu\rho'}K_\nu+g_{\rho'\nu}K_\mu-2g_{\nu\mu}K_{\rho'}\right]\\
    =2K^2\delta_{\rho'}^\rho+(4D-6)K^\rho K_{\rho'},
    \end{split}
\end{align}
where $D=g^{\mu}_\mu$ is the space-time dimension.
With $P^2=K^2=0$, we obtain
\begin{align}
    |\mathcal M|^2&\propto K^\rho K_{\rho'}P^{\tilde\rho}P_{\tilde\rho'}G_{\rho\tilde\rho}(Q)\bar{G}^{\rho'\tilde\rho'}(Q) \nonumber\\
    &\qquad\times\left(1 + \mathcal O\left(\frac{Q}{K},\frac{Q}{P}\right)\right)\\
    &=|(K+K')^\rho (P+P')^{\tilde\rho}G_{\rho\tilde\rho}(Q)|^2.\nonumber\\
    &\qquad\times\left(1 + \mathcal O\left(\frac{Q}{K},\frac{Q}{P}\right)\right)
    \label{eq:leadingorderinQgluon}
\end{align}
Again, to the lowest order in $Q$, this is the same as the scalar quark result \eqref{eq:scalar_quark_result}.

\section{Validitiy of screening prescription\label{app:validity_screening}}

\begin{figure*}
    \includegraphics[width=0.95\linewidth]{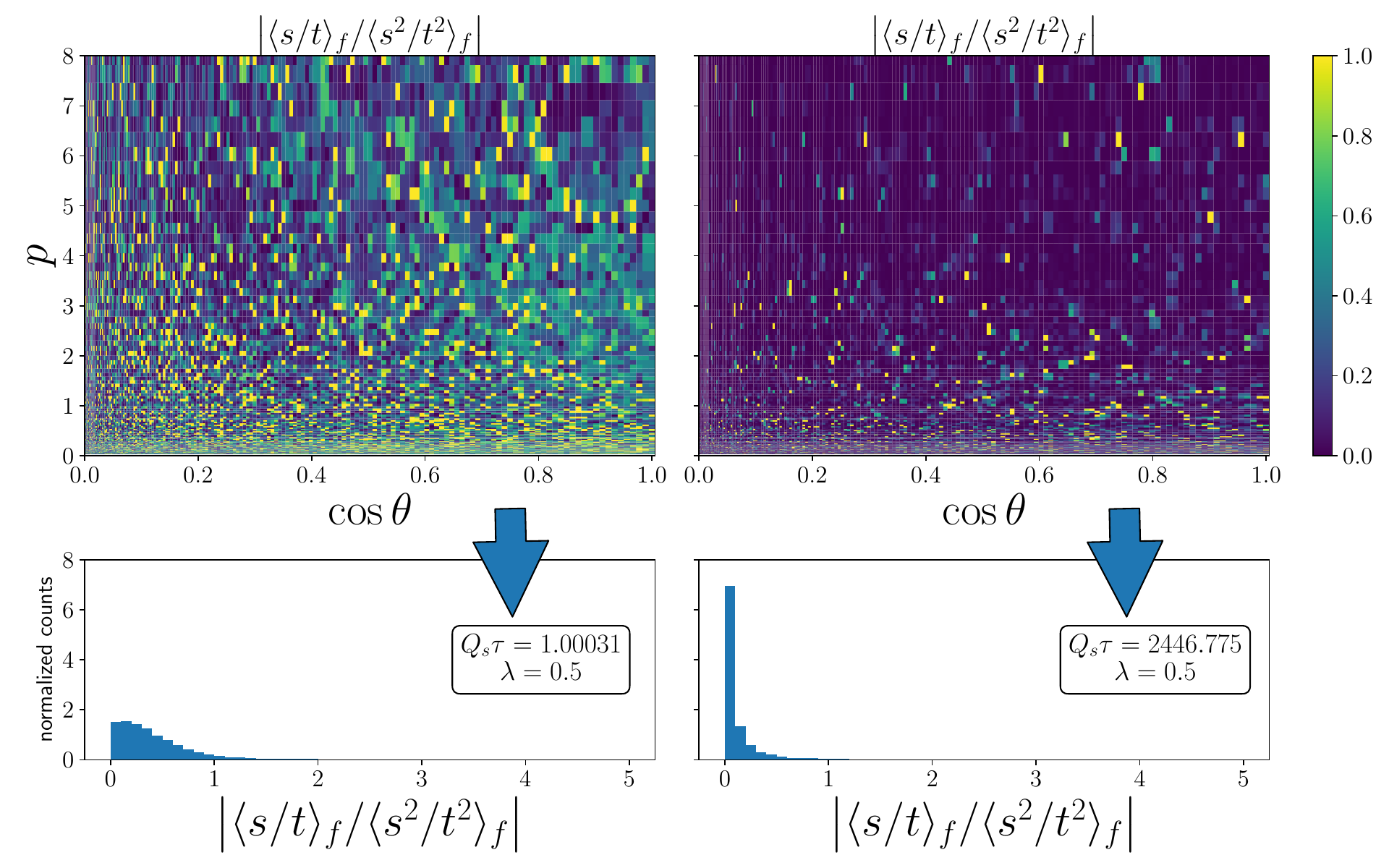}
    \includegraphics[width=0.95\linewidth]{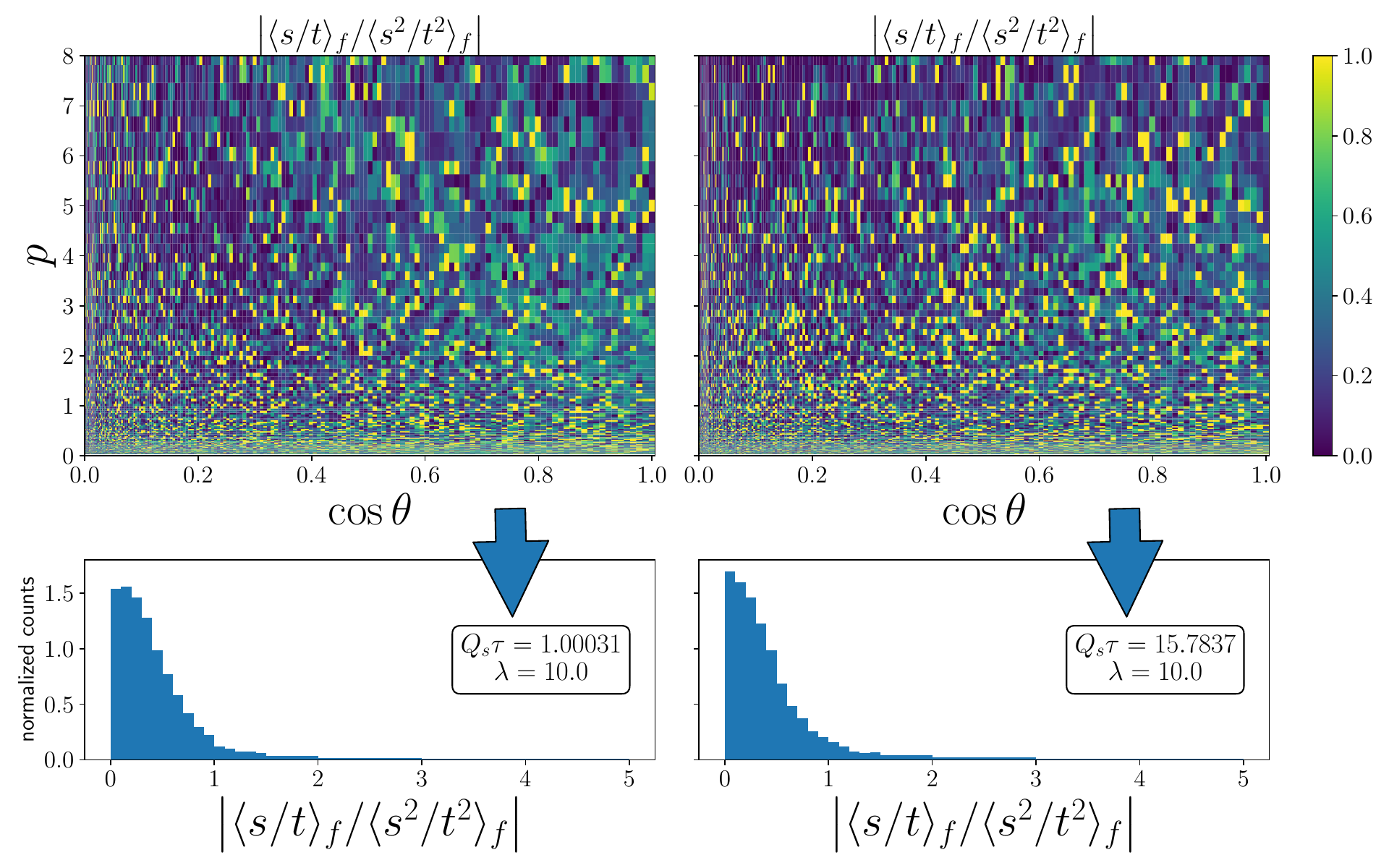}
    \caption{Ratio of changes of the distribution function due to elastic collisions with only $s/t$ as matrix element over $s^2/t^2$, denoted $\left|{\langle s/t\rangle_f}/{\langle s^2/t^2\rangle_f}\right|$, and defined in Eq.~\eqref{eq:rate-of-change-dist-funct-el-col}. Each of the four panels 
    shows $\left|{\langle s/t\rangle_f}/{\langle s^2/t^2\rangle_f}\right|$ as a function of the momentum $\vb p$, parametrized by its length $p$ and angle $\theta$ with respect to the beam axis, and followed by a weighted histogram below it (marked by a blue arrow). 
    The left column corresponds to the initial time and the right column to a later time indicated in the label for the couplings $\lambda=0.5$ {\em (top)} and $\lambda=10$ {\em (bottom)}.
    }
    \label{fig:change_05_10}
\end{figure*}

In this appendix, we demonstrate that $|t|\ll |s|\sim |u|$ is the dominant regime for elastic processes in the collision term. This implies an agreement among the different screening prescriptions that we discussed in \eq \eqref{eq:different_regularization_matrix_element} valid for $|t|/s \ll 1$ and $|t/u| \ll 1$.

To show this, we take a typical distribution function $f(\vb p)$ during the bottom-up process and assess the value of $s/t$ and $s^2/t^2$ when integrated over phase-space and the statistical factors. In particular, 
we perform kinetic theory simulations using the matrix element \eqref{eq:usual_screened_matrix_element}, and use the distribution function at two distinct times.
Effectively, we assess
the contribution to the collision term from the small $|t|\ll s$ contributions,
\begin{align}
    &\left|\frac{\left.\pdv[f(\vb p)]{\tau}\right|_{\text{el. coll. with $s/t$}}}{\left.\pdv[f(\vb p)]{\tau}\right|_{\text{el. coll. with $s^2/t^2$}}}\right| =\left|
        \frac{\int\dd\Gamma\frac{s}{t}\, F^{2 \leftrightarrow 2}}{\int\dd\Gamma\frac{s^2}{t^2}\, F^{2 \leftrightarrow 2}}
    \right|=\left|\frac{\langle s/t\rangle_f}{\langle s^2/t^2\rangle_f}\right|.\label{eq:rate-of-change-dist-funct-el-col}
\end{align}
Here, we introduced the short notation $\langle \dots\rangle_f$ to denote the phase-space average including the distribution functions $F^{2 \leftrightarrow 2} = f_pf_k(1+f_{p'})(1+f_{k'})-f_{p'}f_{k'}(1+f_{p})(1+f_{k})$. Note that the result still depends on the vector $\vb p$, which is the argument of the initial distribution function.

We show these results in \fig \ref{fig:change_05_10} 
for $\lambda=0.5$ and $\lambda=10$ at two distinct time-steps in our simulation. We observe that the values of $|\langle s/t\rangle/\langle s^2/t^2\rangle|$ fluctuate depending on the momentum bin, coupling, and time. However, when plotted within a histogram (indicated by blue arrows), they are consistantly found to be dominated by values below $1$. Therefore, we conclude that indeed $\langle s/t\rangle$ terms are suppressed as compared to $\langle s^2/t^2\rangle$ terms, and the results with the different Debye-like screening prescriptions coincide
not only for weak coupling.

\section{Adaptive step size\label{app:adative_stepsize}}

The philosophy of the adaptive timestep is to set it in such a way that the relative change of specific observables is smaller than a predefined constant $\epsgoal$. This is similar to the stepsize proposed in Ref.~\cite{Du:2020dvp}, but we go beyond linearization and calculate the exact change in the considered observables. We find that this approach is needed to stabilize simulations with isoHTL screening.

To be more concrete, in every timestep $t_{k+1}=t_{k}+\Delta t$, we change the stored values of $n_{ij}$ (defined in \eqref{eq:def_nij}) such that
\begin{align}
	n_{ij}(t_{k+1})=n_{ij}(t_{k})+\Delta t \, c_{ij},
\end{align}
and $c_{ij}$ are numerical estimates (from Monte Carlo integrals) of the sum of all collision terms in Eq.~\eqref{eq:boltzmann_equation}.

Due to the discretization choice \eqref{eq:def_nij}, several quantities can be easily calculated,
\begin{subequations}
\begin{align}
    \varepsilon&=\int\frac{\dd[3]{\vb p}}{(2\pi)^3}p f(\vb p)=\sum_{ij} \frac{n_{ij} p}{\lambda}\\
	n&=\int\frac{\dd[3]{\vb p}}{(2\pi)^3} f(\vb p)=\sum_{ij} \frac{n_{ij} }{\lambda}\\
	n\langle p_z^2\rangle&=\int\frac{\dd[3]{\vb p}}{(2\pi)^3}p^2\cos^2\theta f(\vb p)=\sum_{ij} \frac{n_{ij}}{\lambda} p^2\cos^2\theta\\
	n\langle f\rangle &=\int\frac{\dd[3]{\vb p}}{(2\pi)^3}f^2(\vb p)=\sum_{ij} \frac{n_{ij}^2}{\lambda^2}\frac{(2\pi)^3}{\Delta V_i^p\Delta V_j^\theta p_i^2 4\pi}\\
	n\langle pf\rangle&=\int\frac{\dd[3]{\vb p}}{(2\pi)^3}pf^2(\vb p)=\sum_{ij} \frac{n_{ij}^2}{\lambda^2} p_i\frac{(2\pi)^3}{\Delta V_i^p\Delta V_j^\theta p_i^2 4\pi}\\
	m^2&=2\lambda\int\frac{\dd[3]{\vb p}}{(2\pi)^3} \frac{f(\vb p)}{p}=2\sum_{ij} \frac{n_{ij}} {p_i}\,.
\end{align}
\end{subequations}
Note that except for $n\langle f\rangle$ and $n\langle pf\rangle$, these quantities are linear in $n_{ij}$.
For a given estimate of the collision terms $c_{ij}$, we can accurately predict the change in these observables in a single timestep and adjust the timestep such that the relative change in these observables does not exceed the parameter $\epsgoal$.

For quantities that are linear in $n_{ij}$ (e.g., $\varepsilon$, $m^2$, $n$, $n\langle p_z^2\rangle$) we obtain their exact change per unit time by replacing $n_{ij}\to c_{ij}$. At the example of the energy density $\varepsilon$,
\begin{align}
	\delta\varepsilon:=\frac{\Delta \varepsilon}{\Delta t}&=\sum_{ij} c_{ij} p_i.
\end{align}
With the relative change  
	$\frac{\Delta \varepsilon}{\varepsilon}=\frac{\delta\varepsilon \Delta t}{\varepsilon},$
the new time step can be set according to
\begin{align}
	\Delta t = \frac{\epsgoal}{\left|\frac{\delta\varepsilon}{\varepsilon}\right|}\, .
\end{align}

For the quadratic quantities (e.g., $n\langle f\rangle$ and $n\langle pf\rangle$), the linear approximation (at the example of $\langle f\rangle$)
\begin{align}
	\Delta (n\langle f\rangle) \approx \sum_{ij}2 n_{ij} c_{ij} V_{ij}
\end{align}
with 
\begin{align}
	V_{ij}=\frac{(2\pi)^3}{\Delta V_i^p\Delta V_j^\theta p_i^2 4\pi}
\end{align}
is only a crude estimate of the change and we find that using the correct change leads to much better stabilization. The exact change $\Delta \langle f\rangle = \langle f\rangle_{i+1} - \langle f\rangle_i$ is obtained via
\begin{align}
	\Delta \langle f\rangle &=\sum_{ij}V_{ij}\left( (n_{ij}+c_{ij}\Delta t)^2- n_{ij}^2\right)\\
	&=\sum_{ij}V_{ij}\left(2n_{ij}c_{ij}\Delta t + c_{ij}^2(\Delta t)^2\right).
\end{align}
Enforcing a maximum relative change $|\Delta \langle f\rangle/\langle f\rangle| \leq  \epsgoal$ 
leads to a quadratic equation in $\Delta t$,
\begin{align}
	|a\Delta t+b\Delta t^2|=\epsgoal,\label{eq:quadratic_equation}
\end{align}
with $b > 0$ and
\begin{align}
	a=\frac{\sum_{ij}V_{ij}2 n_{ij}c_{ij}}{\langle f\rangle}, && b = \frac{\sum_{ij}V_{ij}(c_{ij})^2}{\langle f \rangle}.
\end{align}

To solve this quadratic equation for the most restrictive $\Delta t$, we consider two cases:

\begin{enumerate}
    \item \emph{ $a > 0$}:
	
	Here, $a\Delta t+b \Delta t^2$ monotonically increases and is always positive, 
 thus the solution is given by
\begin{align}
	\Delta t = -\frac{a}{2b}+\sqrt{\frac{a^2}{4b^2}+\frac{\epsgoal}{b}}.
\end{align}

\begin{figure}[t!]
    \centering
    \includegraphics[width=\linewidth]{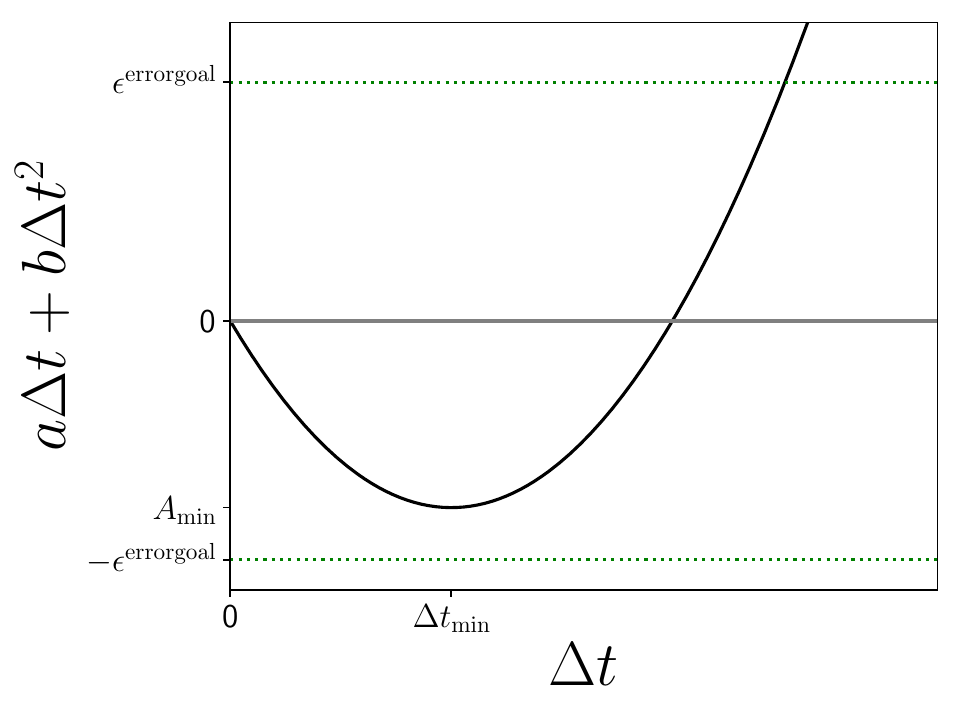}
    \caption{Sketch of the quadratic equation \eqref{eq:quadratic_equation} for $a<0$.}
    \label{fig:parabola}
\end{figure}

\item For \emph{ $a < 0$},
the parabola is sketched in Fig.~\ref{fig:parabola}. Its minimum is at $\Delta t_{\mathrm{min}}=-a/(2b)$, with the minimum value $A_{\mathrm{min}}=-a^2/(4b)$.

When $|A_{\mathrm{min}}|<\epsgoal$ as in Fig.~\ref{fig:parabola}, we need to solve for positive $\epsilon$, which yields the same result as above (again, we need the positive square root, as can be easily seen from the figure)
\begin{align}
		\Delta t = -\frac{a}{2b}+\sqrt{\frac{a^2}{4b^2}+\frac{\epsgoal}{b}}.
\end{align}
For $|A_{\mathrm{min}}|>\epsgoal$, we need to equate to negative $\epsgoal$, i.e., solve
	$a\Delta t+ b\Delta t^2=-\epsgoal,$
which yields
\begin{align}
	\Delta t=-\frac{a}{2b}-\sqrt{\frac{a^2}{4b^2}-\frac{\epsgoal}{b}}.
\end{align}
\end{enumerate}

The new goal for the timestep is then set by determining the minimum of these 
\begin{align}
\Delta t^{\mathrm{goal}}=\min_{i\in \{\varepsilon, n, n\langle p_z^2\rangle, n\langle f\rangle, n\langle pf\rangle, m^2\}}\Delta t_i
\end{align}
and then adjusting it smoothly to the old one via
\begin{align}
	\Delta t^{\mathrm{new}}=\sqrt{(\Delta t^{\mathrm{goal}})^{2\alpha}(\Delta t^{\mathrm{old}})^{2-2\alpha}},
\end{align}
where $\alpha$ is a constant $0 <\alpha < 1$ that 
ensures an interpolation between the previous value and the goal time step. While $\alpha=1$ corresponds to always choosing the goal timestep, $\alpha=0$ corresponds to a fixed time step. Choosing $\alpha$ between these limits prevents large variations of this time step due to statistical fluctuations. In practice, we find a value $\alpha=0.9$ to be convenient in our simulations.

\bibliography{ektmatrixelements}

%%%%%%%%%%%%%%%%%%%%%%%%%%%%%%%%%%%%%%%%%%%%%%%%%%%%%

\end{document}